\documentclass{WileyMSP-template}

\usepackage{cite}
\usepackage{amsmath,amssymb,amsfonts}
\usepackage{algorithmic}
\usepackage{algorithm}
\usepackage{graphicx}
\usepackage{textcomp}
\usepackage{xcolor}
 
\usepackage{url}            % simple URL typesetting
\usepackage{booktabs}       % professional-quality tables
\usepackage{amsfonts}       % blackboard math symbols
\usepackage{nicefrac}       % compact symbols for 1/2, etc.
\usepackage{microtype}      % microtypography
\usepackage{xcolor}         % colors

\usepackage{amssymb} % \mathbb
\usepackage{multicol}
\usepackage{multirow}
\usepackage{pifont}% http://ctan.org/pkg/pifont
\usepackage{makecell}

\begin{document}

\pagestyle{fancy}
\setlength{\headheight}{25pt}
\rhead{\includegraphics[width=2.5cm]{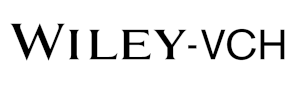}}

\title{MetaGEM: Bottom-Up Reconstruction of Genome-Scale Metabolic Networks via Deep Enzyme–Metabolite Anchoring}

{\sloppy\maketitle\par}

% Author: Please give full first and last names for authors and include * after the name of all corresponding authors

% Dedication

\dedication{}

% Affiliations: Please provide adacemic titles (Prof. or Dr.) for all authors where applicable
\begin{affiliations}
Weiyu Xiao$^{1,2 \dagger}$, Jiangbin Zheng$^{2,3,4 \dagger}$, and Stan Z. Li$^{2 *}$\\[0.7em]

$^{1}$Huazhong University of Science and Technology, Wuhan, China

$^{2}$Westlake University, Hangzhou, China

$^{3}$Zhejiang University, Hangzhou, China

$^{4}$Shanghai Artificial Intelligence Laboratory, Shanghai, China
\end{affiliations}

% Keywords: Please provide a minimum of three and a maximum of seven keywords, separated by commas

% Abstract should be written in the present tense and impersonal style (i.e., avoid we), and be at most 200 words long

\section*{Abstract}

Genome-scale metabolic models (GEMs) are foundational for systems biology and rational chassis design. However, traditional top-down reconstruction heavily relies on sequence homology, leaving vast "metabolic dark matter" unannotated. Conversely, direct bottom-up reconstruction from metabolomics is a classically ill-posed inverse problem plagued by combinatorial explosion and unconstrained network hallucinations. Here, we present MetaGEM, a novel bottom-up paradigm that resolves this dilemma by utilizing enzymes as definitive physical anchors, transforming ambiguous macro-network inference into a precise micro-level molecular pairing task. MetaGEM employs a multimodal dual-tower deep learning architecture, seamlessly integrating profound evolutionary semantics from protein language models with the 3D spatial conformations of metabolites. Crucially, it introduces a contrastive learning framework with hard negative mining to distinguish fine-grained structural nuances, effectively overcoming the prevalent high-similarity interference in metabolic spaces. Extensive benchmarking demonstrates that MetaGEM achieves state-of-the-art performance in enzyme-metabolite interaction prediction (AUROC 0.9701, MCC 0.8033) and exhibits exceptional zero-shot generalization toward evolutionarily distinct, uncharacterized proteins. In system-level applications, MetaGEM-driven pipelines successfully reverse-engineered highly connected, functional GEMs for diverse organisms (\textit{E. coli}, \textit{B. subtilis}, and \textit{P. aeruginosa}). By accurately capturing promiscuous enzymes, it significantly rescues network connectivity and demonstrates superior biological fidelity in \textit{in silico} phenotype microarrays and gene essentiality predictions. By bridging the gap between dynamic metabolomic snapshots and computable metabolic networks, MetaGEM illuminates metabolic dark matter and establishes a rigorous computational foundation for the automated generation of AI-driven virtual cells.

\medskip

\noindent\textbf{Keywords:} Genome-Scale Metabolic Models, Enzyme-Metabolite Interactions, Metabolite-Driven Reconstruction, Deep Contrastive Learning, Metabolic Dark Matter, Promiscuous Enzymes

\section{Introduction}

For a long time, genome-scale metabolic model (GEM) reconstruction has relied heavily on a ``top-down'' genome annotation paradigm \cite{baart2011genome,seaver2021modelseed}: enzyme functions are first inferred through sequence homology alignment \cite{kharchenko2006identifying,machado2018fast}, and then mapped onto biochemical reaction networks. However, this paradigm is now facing an insurmountable bottleneck \cite{mueller2013rapid,dahal2016genome,de2024pan}. When confronted with non-model organisms, extremophiles, and increasingly massive metagenomic datasets, nature presents a vast number of orphan genes and proteins of unknown function \cite{sorokina2014profiling,lobb2015remote}. Owing to the lack of known homologous motifs, conventional metabolic reconstruction methods often fail in the face of such ``metabolic dark matter'' \cite{escudeiro2022functional,palsson2026approaches}, resulting in reconstructed models filled with network gaps that cannot faithfully reflect the true metabolic potential of the organism \cite{hsieh2024comparative}.

Meanwhile, with the rapid development of high-resolution mass spectrometry and metabolomics, we are entering an era in which the mode of data acquisition is undergoing a fundamental transformation \cite{perez2021ultra,zhang2024mass,ali2022single}. Compared with genomics, which represents static potential, metabolomics directly provides a dynamic snapshot of the cellular functional state \cite{gonccalves2021genome,qiu2023small,zhang2025dynamic}, revealing the actual repertoire of small molecules actively circulating within the cell. These readily accessible molecular features, rich in phenotypic information \cite{buergel2022metabolomic,nightingale2024metabolomic}, offer a fundamentally new perspective: can we change course, circumvent the dependence on sequence homology, and directly start from observed metabolites to reconstruct, in a ``bottom-up'' manner, the metabolic network that drives life activities?

Although such ``phenotype-driven reverse engineering'' is conceptually highly attractive, in computational science it constitutes a classic ill-posed inverse problem \cite{engl2009inverse,li2023covrecon,moseley2013error}. Mapping discrete chemical small molecules to a system-level network faces the computational challenge of combinatorial explosion \cite{klamt2002combinatorial}: an individual metabolite is often a hub in a large biochemical network \cite{waller2020compartment}, and its mere presence is insufficient to determine whether it acts as a substrate, a product, or a bypass intermediate \cite{blair2012can}. In the absence of catalysts (enzymes) as biological constraints, simply assembling reactions on the basis of stoichiometry can only generate an illusory network that is thermodynamically feasible but biologically spurious \cite{krumholz2017thermodynamic,sanchez2017improving,yang2023improving}.

\begin{figure*}
    \centering
    \includegraphics[width=0.99\linewidth]{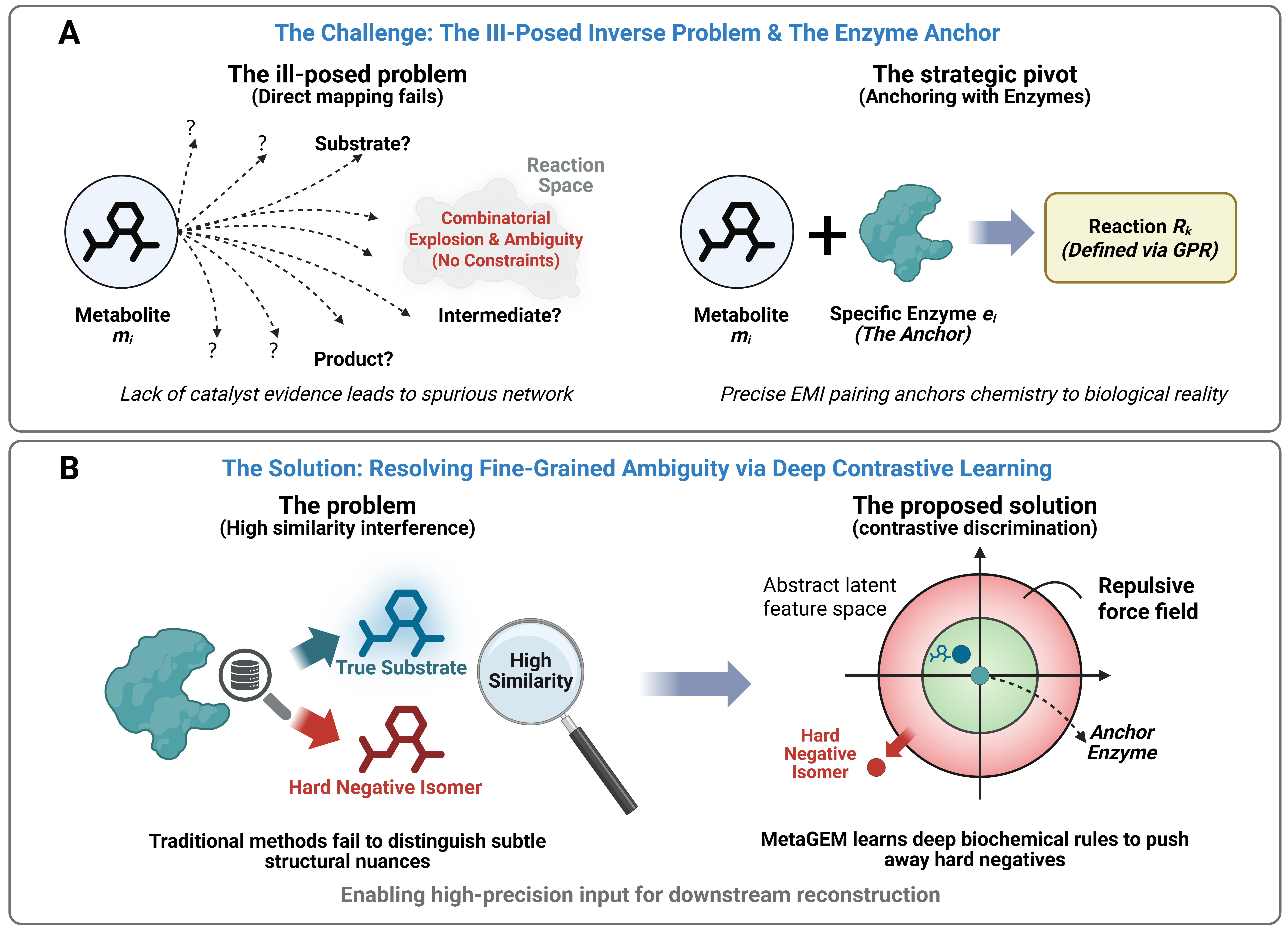}
    \caption{Core challenges of bottom-up metabolic network reconstruction and the conceptual framework of MetaGEM.
(a) Core challenge and strategic fulcrum: resolving the ill-posed inverse problem. The left panel illustrates the ``ill-posed inverse problem'' encountered when directly mapping metabolites into reaction space: the absence of catalytic constraints leads to severe combinatorial explosion and directional ambiguity. The right panel presents the strategic fulcrum of this study: introducing enzymes as the biological anchor. By accurately predicting enzyme--metabolite interactions (EMIs), the ambiguous problem of network inference is transformed into precise physical molecular pairing, thereby enabling the definition of authentic biochemical reactions through GPR rules.
(b) The MetaGEM solution: eliminating fine-grained ambiguity through deep contrastive learning. Faced with highly confusable isomers and structurally similar analogs in metabolic space (high-similarity interference), conventional methods struggle to distinguish them (left). MetaGEM innovatively introduces a contrastive discrimination mechanism (right), constructing a repulsive force field in an abstract latent feature space. By learning deep biochemical rules, it forcibly pushes highly similar hard negatives/isomers away from the anchor enzyme, thereby providing high-precision and highly specific component inputs for downstream macroscopic network reconstruction.}
    \label{fig:metagem_1}
\end{figure*}

To resolve this ill-posed inverse problem, we require a physical constraint capable of anchoring chemical potential to biological entities \cite{sanchez2017improving,li2022deep}. According to the classical gene--protein--reaction (GPR) association rules, enzymes are precisely such a natural hub \cite{ryu2017framework,machado2016stoichiometric,di2021gpruler}. On this basis, we transform the macroscopic problem of network inference into a microscopic molecular pairing task, namely the accurate prediction of enzyme--metabolite interactions (EMIs) \cite{piazza2018map,kroll2023general,kroll2024multimodal,nie2025omniesi} (Figure~\ref{fig:metagem_1}a). If we can accurately match each molecule in the metabolome to its specific catalytic enzyme, we can obtain a well-defined functional enzyme set \cite{kroll2023general}, which can then be seamlessly converted into a high-confidence reaction network according to GPR rules \cite{machado2016stoichiometric}, thereby transforming originally ambiguous network inference into precise physical constraint matching \cite{habibpour2024prediction}.

However, achieving high-fidelity EMI prediction faces a major technical barrier, namely high similarity interference \cite{kroll2023general,salas2024machine}. Metabolic space is populated with structurally highly similar isomers and derivatives that are nearly indistinguishable using conventional chemical fingerprints \cite{du2023fusing,orsi2024one}, yet are catalyzed by entirely different enzymes. To overcome this barrier, this chapter proposes a multimodal predictive framework integrated with deep contrastive learning---MetaGEM (Figure~\ref{fig:metagem_1}b). MetaGEM abandons dependence on shallow sequence features and, from first principles, leverages large-scale pretrained models to extract deep evolutionary semantics of proteins and 3D geometric conformational features of metabolites, respectively. More importantly, we innovatively introduce a contrastive learning mechanism based on hard negative mining. By constructing a repulsive force field in the latent feature space, the model is forced to push away highly similar decoy molecules that would otherwise be incorrectly pulled close, thereby endowing it with the ability to distinguish subtle differences in chemical structure.

In summary, as the metabolite-oriented branch of macroscopic network construction, this chapter proposes an entirely new paradigm for GEM reconstruction. Through MetaGEM, we first accomplish an accurate mapping from a ``metabolite set'' to a ``predicted enzyme set.'' Subsequently, drawing inspiration from mature automated reconstruction strategies, we design a mathematical constraint and inference mechanism based on the predicted enzyme set. This mechanism uses GPR rules to map predicted enzymes to reaction templates, and, by introducing stoichiometric balance and thermodynamic constraints, automatically fills network gaps, ultimately assembling a functionally complete genome-scale metabolic model. To the best of our knowledge, this is the first complete computational paradigm proposed and realized for progressing ``from a metabolite puzzle to a whole-genome model.'' Extensive experiments demonstrate that the MetaGEM-driven reconstruction method significantly outperforms existing mainstream baseline methods across tasks involving multiple model strains, exhibiting substantial potential for resolving the metabolic ``dark matter'' of complex biological systems.

The main contributions of this chapter can be summarized in the following four aspects:

$\bullet$ \textbf{Pioneering a new bottom-up paradigm for inverse reconstruction:} We establish, for the first time, a computational route for the Metabolite2GEM task that does not rely on conventional sequence annotation, but instead directly starts from metabolomics data and reconstructs genome-scale metabolic networks using enzymes as functional anchors, thereby providing a novel theoretical framework for the analysis of complex biological systems.

$\bullet$ \textbf{Proposing the MetaGEM framework for fine-grained biochemical matching:} To address the challenge posed by high chemical similarity in metabolite space, we innovatively introduce a contrastive learning mechanism based on hard negative mining, effectively overcoming the bottleneck of conventional methods in distinguishing isomers and analogs and enabling precise prediction at the level of first principles.

$\bullet$ \textbf{Constructing an end-to-end automated GEM reconstruction and inference system based on mathematical constraints:} Taking the predicted discrete enzyme profile as input, the system automatically assembles a functionally complete and computable metabolic model by integrating GPR association rules, stoichiometric balance, and thermodynamic constraint mechanisms, thus achieving a seamless transition from molecular evidence to a system-level model.

$\bullet$ \textbf{Establishing a rigorous two-level evaluation benchmark and demonstrating SOTA performance on downstream tasks:} We propose a comprehensive evaluation framework covering both microscopic interaction prediction and macroscopic biological functional validation (e.g., \textit{in silico} growth simulation and gene essentiality prediction). Extensive experiments show that, even in the absence of prior homology information, the model still demonstrates a strong capability to uncover potential metabolic pathways and unknown enzyme functions.

\section{Related Work}

\subsection{Conventional GEM Reconstruction: From Homology Annotation to Automated Pipelines}
Genome-scale metabolic model (GEM) reconstruction has long been built upon the classical workflow of ``gene annotation $\rightarrow$ reaction mapping $\rightarrow$ gap filling.'' Representative frameworks include automated tools based on homology alignment and template library expansion, as well as model governance systems emphasizing standardized quality control and reproducible evaluation\cite{thiele2010protocol,seaver2021modelseed,lieven2020memote}. These methods have promoted GEMs from manual expert curation to scalable production, and have achieved stable performance in model strains\cite{baart2011genome,machado2018fast}. However, their core premise is that ``sufficient homologous evidence is available.'' When the research object extends to non-model organisms, extremophiles, or metagenomic contexts, orphan genes and low-homology sequences are pervasive, and traditional workflows often suffer from the dual problems of reaction under-recall, network disconnections, and excessive gap-filling\cite{mueller2013rapid,dahal2016genome,de2024pan,escudeiro2022functional}.

\subsection{Enzyme--Substrate/Enzyme--Metabolite Prediction: From Rule-Driven Methods to Deep Learning}
To break through the ``annotation-first'' paradigm, researchers have gradually shifted their focus to more direct inference of molecular-level relationships, namely enzyme--substrate or enzyme--metabolite interaction (ESI/EMI) prediction. Early methods mostly relied on reaction rules, molecular fingerprints, and manual feature engineering, which could provide interpretable results in constrained chemical spaces, but had limited generalization ability in high-dimensional and noisy scenarios\cite{du2023fusing,wang2020deep}. In recent years, deep learning models have significantly raised the upper bound of this task: through collaborative modeling of protein sequence language models and molecular representation networks, methods such as ESP, VIPER, and OmniESI have continuously improved discriminative performance on unseen samples\cite{kroll2023general,Campbell2024.06.21.599972,nie2025omniesi}. At the same time, multimodal fusion and pretrained paradigms have further demonstrated the complementary value of ``evolutionary semantics + geometric conformation'' in recognizing catalytic specificity\cite{kroll2024multimodal,lin2022language,zhou2023unimol}.

\subsection{Key Bottlenecks: Homology Leakage and High-Similarity Interference}
Despite the rapid progress of existing ESI/EMI methods, two systematic pain points remain in the literature. First, homology leakage at the data-splitting level can substantially overestimate model generalization ability: if the training and test sets share highly similar enzyme families, the model may degenerate into a ``sequence memorizer'' rather than a true learner of biochemical rules\cite{fu2012cd,kroll2023general}. Second, the widespread presence of isomers and neighboring derivatives in metabolic space causes high-similarity interference; traditional two-dimensional fingerprints or weakly supervised representations often struggle to distinguish fine-grained structural differences, leading to increased false positives and their amplification into erroneous network edges during downstream reconstruction\cite{salas2024machine,tahil2024stereoisomers,zhu2024metapredictor,orsi2024one}.

\subsection{Positioning of This Study: Connecting Molecular Prediction and System Reconstruction with EMI as the Anchor}
Compared with existing studies, the goal of this work is not merely to improve a single binary classification metric, but to explicitly embed EMI prediction into the closed loop of GEM reconstruction. We regard the ``enzyme'' as the physical anchor connecting metabolite observations and reaction network assembly: first, molecular pairing fidelity is improved through homology-aware splitting and hard-negative contrastive learning; then, system-level assembly is completed based on GPR rules, stoichiometric constraints, and thermodynamic constraints\cite{machado2016stoichiometric,di2021gpruler,sanchez2017improving}. Therefore, this work lies at the intersection of two research lines: it builds upon the enzyme--metabolite learning paradigm and extends toward the constraint-driven network reconstruction framework, forming an integrated route ``from metabolite puzzle to whole-genome model.''

\section{Methods}

\begin{figure*}
    \centering
    \includegraphics[width=0.92\linewidth]{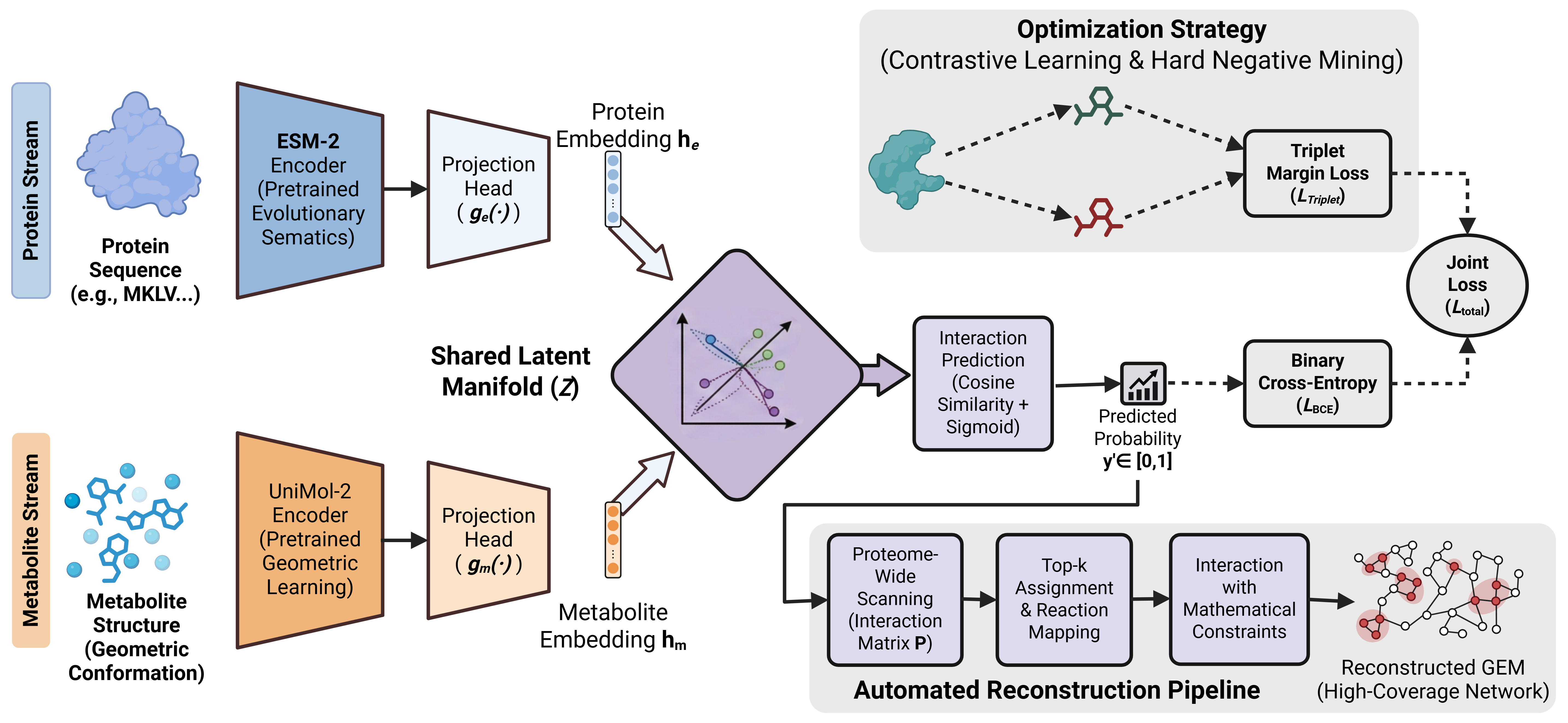}
    \caption{\textbf{MetaGEM computational framework: from deep dual-tower feature alignment to system-level network reconstruction.} The figure shows the complete end-to-end workflow. First, the dual-tower multimodal architecture (left) uses pretrained ESM-2 and UniMol-2 encoders to extract the deep evolutionary semantics of proteins ($\mathbf{h}_e$) and the three-dimensional geometric conformations of metabolites ($\mathbf{h}_m$), respectively, and maps them into a shared latent manifold space ($\mathcal{Z}$) through nonlinear projection heads for alignment and interaction prediction. Second, the optimization strategy (upper right) uses a joint loss function ($\mathcal{L}_{total}$) for model training, which combines binary cross-entropy for classification ($\mathcal{L}_{BCE}$) with triplet margin loss based on hard negative mining ($\mathcal{L}_{Triplet}$) to enhance the model's fine-grained discriminative ability toward structural analogs. Finally, the automated reconstruction pipeline (lower right) uses the trained model to perform whole-proteome scanning (generating the interaction matrix $\mathbf{P}$), extracts high-confidence drafts through Top-k assignment and reaction mapping, and assembles them into a high-coverage functional genome-scale metabolic model (GEM) under strict mathematical constraints.}
    \label{fig:2}
\end{figure*}

\subsection{Preliminaries and Problem Formulation}
We abstract the core challenge of genome-scale metabolic model (GEM) reconstruction as a deep \textbf{link prediction} problem, aiming to construct a high-fidelity mapping between the enzyme sequence space $\mathcal{E}$ and the metabolite chemical space $\mathcal{M}$.

Formally, let the enzyme set be $\mathcal{E} = \{e_i\}_{i=1}^{N_e}$ and the metabolite set be $\mathcal{M} = \{m_j\}_{j=1}^{N_m}$. We model enzyme--metabolite interactions (EMIs) as a binary relation matrix $Y \in \{0, 1\}^{N_e \times N_m}$, where $y_{ij}=1$ indicates that enzyme $e_i$ catalyzes a reaction involving metabolite $m_j$ as a specific substrate. Our goal is to learn a parameterized neural function $f_\theta: \mathcal{E} \times \mathcal{M} \rightarrow [0, 1]$ to approximate the posterior interaction probability $P(y_{ij}=1 \mid e_i, m_j)$.

\subsection{Dataset Construction: A Homology-Aware Protocol}
To construct high-confidence supervision signals and mitigate the bias of traditional sequence alignment methods, we strictly curated and integrated data from the BiGG, Rhea, UniProt, and ChEBI databases.

\textbf{Positive Set Construction}: We define the positive sample set as $\mathcal{D}^+ = \{(e, m) \mid \text{reaction}(e) \ni m\}$, and treat compounds on both sides of reversible reactions as potential substrates. The final MetaGEM-DB dataset contains 330,432 interaction entries, covering evidence from experimentally validated data and phylogenetic inference.

\textbf{Leakage-Proof Data Splitting}: A critical flaw in many existing methods is the use of random splitting, which leads to severe homology leakage and causes the model to degenerate into a simple sequence memorizer. To prevent this, we use CD-Hit to cluster enzyme sequences, defined as $\mathcal{C} = \text{Cluster}(\mathcal{E}, \delta)$. We split the dataset based on these clusters to ensure a strict mutual exclusivity relationship between the training set $\mathcal{E}_{train}$ and the test set $\mathcal{E}_{test}$:
\begin{equation}
\forall e_a \in \mathcal{E}_{train}, \forall e_b \in \mathcal{E}_{test}, \quad \text{sim}_{seq}(e_a, e_b) < \delta
\end{equation}
This cluster-based splitting strategy (8:2 ratio) forces the model to learn universal, first-principles biochemical rules of substrate recognition, rather than memorizing specific sequence motifs.

\subsection{MetaGEM Architecture: A Multimodal First-Principles Approach}
To go beyond the limitations of homology-based inference, MetaGEM adopts an end-to-end deep multimodal computational framework (Figure~\ref{fig:2}). Its core dual-tower architecture (left side of Figure~\ref{fig:2}) is designed to project heterogeneous biological entities into a shared latent manifold. Our goal is to align the evolutionary semantics of proteins with the geometric conformations of small molecules, leveraging state-of-the-art pretrained models to capture fundamental biochemical principles.

\subsubsection{Protein Encoder: Evolutionary Semantic Extraction}
To capture the deep functional semantics of enzymes that cannot be reflected by simple sequence similarity, we adopt ESM-2 (Evolutionary Scale Modeling)\cite{lin2022language}, which is pretrained on genomic data with Masked Language Modeling and learns rich panoramic representations of proteins. Given an enzyme sequence $S$, we extract its contextual embeddings and obtain the global representation $\mathbf{h}_e \in \mathbb{R}^{d_p}$ through mean pooling:
\begin{equation}
\mathbf{h}_e = \frac{1}{L} \sum_{k=1}^{L} \text{ESM}_{\phi}(S)_k
\end{equation}
This representation encodes implicit structural and functional properties derived from billions of years of evolutionary pressure.

\subsubsection{Metabolite Encoder: Geometric Conformation Learning}
Biological activity is highly dependent on 3D spatial conformation, especially when distinguishing stereoisomers or structural analogs. We employ the Uni-Mol2\cite{zhou2023unimol} Transformer architecture, which explicitly models molecular geometry through pretraining tasks such as denoising atomic coordinate prediction. For a metabolite graph $G$, the model outputs an embedding vector $\mathbf{h}_m \in \mathbb{R}^{d_m}$ containing rich geometric priors, which is crucial for specific binding recognition.

\subsubsection{Multimodal Alignment and Inference}
To align these two heterogeneous feature spaces, we introduce nonlinear projection heads $g_e(\cdot)$ and $g_m(\cdot)$ to map $\mathbf{h}_e$ and $\mathbf{h}_m$ into a metric space $\mathcal{Z}$ of dimension $d$, obtaining normalized representations $\mathbf{z}_{e_i}$ and $\mathbf{z}_{m_j}$, respectively. In this space, the interaction probability is computed through scaled cosine similarity with a learnable temperature parameter $\tau$ and bias $b$:
\begin{equation}
\mathbf{z}_{e_i} = \frac{g_e(\mathbf{h}_{e_i})}{\|g_e(\mathbf{h}_{e_i})\|}, \quad \mathbf{z}_{m_j} = \frac{g_m(\mathbf{h}_{m_j})}{\|g_m(\mathbf{h}_{m_j})\|}
\end{equation}
\begin{equation}
\hat{y}_{ij} = \sigma \left( \frac{\mathbf{z}_{e_i}^\top \cdot \mathbf{z}_{m_j}}{\tau} + b \right)
\end{equation}
where $\sigma(\cdot)$ is the Sigmoid activation function, and $\tau$ is used to regulate the smoothness of the logistic distribution, ensuring that the model can break through the value-range limitation of standard cosine similarity and output prediction probabilities with extremely high confidence. This design ensures that the final prediction is based on deep semantic alignment between the functional potential of proteins and the structural compatibility of metabolites.

\subsection{Optimization: Joint Classification and Contrastive Learning}
The core challenge of EMI prediction is the problem of ``high-similarity interference,'' where chemically similar molecules often have entirely different biological roles. To address this issue, we propose a joint training framework that integrates the classification task and the contrastive learning objective (upper right of Figure~\ref{fig:2}). The classification module directly optimizes prediction accuracy, whereas the contrastive module serves as a key regularization term to enhance the separability of fine-grained representations in latent space.

\subsubsection{Hard Negative Mining Strategy}
Because the number of non-interacting pairs far exceeds that of interacting pairs, and to improve the model's ability to discriminate ``decoy'' molecules (such as isomers), we implement a \textbf{hard negative mining} strategy. For each positive sample pair $(e_i, m_i^+)$, we sample a negative instance $m_i^-$ that has high chemical similarity to the true substrate $m_i^+$ (measured by the Tanimoto coefficient):
\begin{equation}
m_i^- \sim \{m \in \mathcal{M} \setminus \mathcal{N}(e_i) \mid \text{sim}_{chem}(m, m_i^+) > \epsilon \}
\end{equation}
This ensures that the model learns robust boundary features by focusing on ``hard cases'' located near the decision boundary. During training, the ratio of positive samples to hard negative samples is set to 1:30.

\subsubsection{Joint Loss Function}
The model is trained end-to-end by minimizing the joint loss function $\mathcal{L}_{total}$:
\begin{equation}
\mathcal{L}_{total} = \mathcal{L}_{BCE} + \lambda_{trip} \mathcal{L}_{Triplet}
\end{equation}

\textbf{Classification Loss ($\mathcal{L}_{BCE}$)}: We use binary cross-entropy loss to ensure that the model has the basic discriminative performance for interaction probability, guiding the overall alignment direction.
\begin{equation}
\mathcal{L}_{BCE} = - \mathbb{E}_{(e,m,y) \sim \mathcal{D}} [y \log \hat{y} + (1-y) \log (1-\hat{y})]
\end{equation}

\textbf{Contrastive Regularization ($\mathcal{L}_{Triplet}$)}: We employ triplet margin loss to explicitly optimize the topological structure of the feature space. We define the distance metric in latent space as $d(e, m) = 1 - \mathbf{z}_e^\top \cdot \mathbf{z}_m$. By minimizing the distance between the anchor enzyme and its true substrate while maximizing the distance to the hard negative sample, this loss term sharpens the decision boundary:
\begin{equation}
\mathcal{L}_{Triplet} = \mathbb{E}_{(e_i, m_i^+, m_i^-)} \left[ \max(0, d(e_i, m_i^+) - d(e_i, m_i^-) + \alpha) \right]
\end{equation}
where $\alpha$ is a preset margin hyperparameter, which forces hard negative samples with slight structural differences to be pushed beyond the decision boundary, thereby endowing the model with first-principles-level fine-grained discriminative power.

\subsection{Automated Reconstruction Pipeline: From Molecular Evidence to System-Level Models}
Leveraging the high-precision predictions of MetaGEM, we propose a novel end-to-end reconstruction pipeline (lower right of Figure~\ref{fig:2}). This pipeline anchors the observed metabolic potential to specific genetic entities and assembles a functionally complete GEM using strict mathematical constraints.

1. \textbf{Proteome-Wide Scoring}: Given the observed metabolome $\mathcal{M}_{obs}$ of a target organism and a candidate protein library, the trained MetaGEM is used to compute the interaction probability matrix $\mathbf{P}$ for all potential enzyme--metabolite pairs.

2. \textbf{Diversity-Penalized Top-k Enzyme Assignment}: Traditional hard assignment (e.g., directly selecting Top-1) faces a severe ``supply--demand imbalance'' in practical system construction: because the number of observed metabolite species is far smaller than the size of the candidate enzyme library, Top-1 matching leads to an extracted enzyme set that is too small, and after deduplication is insufficient to support the global connectivity of the network. However, simply adopting a widened-window Top-k strategy causes a severe ``confidence monopoly'' phenomenon---that is, multiple different metabolites tend to cluster onto the same high-confidence ``hub enzyme,'' causing other enzymes with biological relevance to be systematically ignored.
To break this monopoly and maximize biochemical diversity, we introduce a dynamic ``diversity penalty'' during iterative extraction. For metabolite $m$ and enzyme $e$, the actual matching score is calculated as follows:
\begin{equation}
\text{Score}(m,e) = \text{OriginalScore}(m,e) - \lambda \times \text{Count}(e)
\end{equation}
where $\text{OriginalScore}(m,e)$ is the initial predicted probability output by the MetaGEM model; $\text{Count}(e)$ denotes the number of times that the enzyme has already been selected by other metabolites in the current round; and $\lambda$ is the diversity penalty coefficient (set to $\lambda=0.2$ in empirical studies). Under this penalty mechanism, we perform Top-2 sampling for each metabolite. This dynamic decay strategy forces the model to explore a broader candidate enzyme space beyond local optima, effectively enriching the enzyme diversity required for downstream network construction while ensuring high-confidence molecular pairing. The selected deduplicated enzyme set finally constitutes the predicted functional proteome $\mathcal{E}_{pred}$.

3. \textbf{Reaction Mapping \& Draft Construction}: The predicted enzyme set $\mathcal{E}_{pred}$ is used as input and mapped to BiGG reaction templates. We use an established protocol to convert enzyme sequences into specific biochemical reactions and construct the initial draft metabolic network.

\subsection{Mathematical Modeling and Constraint Optimization of the Reconstructed Network}
After the initial draft has been constructed, the network is transformed into a computable mathematical matrix. We apply a series of constraint-based optimization algorithms to define gene--protein--reaction associations, simulate metabolic fluxes, and refine the model to ensure biological feasibility.

\subsubsection{Construction of Gene--Protein--Reaction (GPR) Logic Gates}
To bridge the gap between predicted genotype and phenotype, MetaGEM transforms discrete sequence mappings into a rigorous Boolean logic network. The relationships are described as follows:

1) Enzyme complex (AND relationship): $Reaction \iff (Gene_A \land Gene_B)$

2) Isozyme (OR relationship): $Reaction \iff (Gene_C \lor Gene_D)$

For each reaction $j$, its activity state $r_j$ is determined by the expression states of its associated genes: $r_j = f_{GPR}(g_1, g_2, ..., g_k), \quad g_i \in \{0, 1\}$. This transformation establishes a rigorous computational basis for subsequent gene knockout and essentiality simulations.

\subsubsection{Stoichiometric Matrix Construction and Flux Balance Analysis (FBA)}
The core representation of the reconstructed network is the stoichiometric matrix $\mathbf{S}$ of dimension $m \times n$, where $S_{ij}$ represents the coefficient of metabolite $i$ in reaction $j$. Under the quasi-steady-state assumption (QSSA), namely that the dynamic change of metabolites satisfies $\frac{d\mathbf{x}}{dt} = \mathbf{S} \cdot \mathbf{v} = \mathbf{0}$, we use linear programming (LP) to solve the flux distribution $\mathbf{v}$ so as to maximize the biomass production objective $Z$:
\begin{equation}
\begin{aligned} 
\text{maximize} \quad & Z = \mathbf{c}^T \cdot \mathbf{v} \\ 
\text{subject to} \quad & \mathbf{S} \cdot \mathbf{v} = \mathbf{0} \\ 
& v_{j, \text{min}} \leq v_j \leq v_{j, \text{max}} 
\end{aligned}
\end{equation}

\subsubsection{Gap Filling Based on Mixed-Integer Linear Programming (MILP)}
To address the inevitable connectivity gaps in the initial draft model, we use a mixed-integer linear programming (MILP) algorithm to restore network connectivity by selectively importing the minimal necessary reaction set from a universal database:
\begin{equation}
\begin{aligned} 
\text{minimize} \quad & \sum_{j \in \text{Univ}} \alpha_j \cdot y_j \\ 
\text{subject to} \quad & \mathbf{S}_{\text{draft}} \cdot \mathbf{v}_{\text{draft}} + \mathbf{S}_{\text{univ}} \cdot \mathbf{v}_{\text{univ}} = \mathbf{0} \\ 
& v_{\text{biomass}} \geq v_{\text{target}} \\ 
& v_{j, \text{min}} \cdot y_j \leq v_{\text{univ}, j} \leq v_{j, \text{max}} \cdot y_j, \quad \forall j \in \text{Univ} \\
& y_j \in \{0, 1\} 
\end{aligned}
\end{equation}
where $y_j$ is a binary indicator variable that controls the activation state of reaction $j$ in the universal database. The Big-M constraints ensure that the flux $v_{\text{univ}, j}$ is allowed to lie within the thermodynamic bounds only when the reaction is selected ($y_j = 1$). $\alpha_j$ is a penalty weight inversely proportional to the MetaGEM prediction probability:
\begin{equation}
    \alpha_j = 1 - \hat{P}(Reaction_j \mid \mathcal{E}_{pred}, \mathcal{M}_{obs})
\end{equation}

This weighting mechanism based on prediction confidence ensures that the system preferentially selects biochemical reactions supported by strong physicochemical evidence, thereby making minimal and biologically grounded corrections to the network.

\section{Experimental Setup}

\subsection{Data Sources and Benchmark Dataset Construction}
To construct a standardized benchmark for high-fidelity enzyme--metabolite interaction (EMI) prediction, this study deeply integrated multi-omics and three-dimensional structural data from the BiGG\cite{schellenberger2010bigg}, Rhea\cite{bansal2022rhea}, UniProt\cite{uniprot2019uniprot}, and ChEBI\cite{gaulton2012chembl} databases to build a comprehensive dataset, MetaGEM-DB.

The specific data curation workflow is as follows: first, all reaction entries from existing genome-scale metabolic models (GEMs) were extracted from the BiGG database and systematically mapped to the standard reaction identifiers in the Rhea database. Subsequently, using Rhea IDs as the bridge for cross-database mapping, the 3D structural coordinates and chemical descriptors of the corresponding metabolites were accurately extracted from ChEBI, while the standard protein sequences of the corresponding catalytic enzymes were synchronously obtained from UniProt.

For the definition of positive samples (interacting pairs), this study strictly followed the directional biochemical reaction annotations provided by Rhea. Compounds on the consumption side (left side) of the reaction equation were explicitly annotated as the specific substrates of the corresponding enzymes; for reversible reactions, compounds on both sides of the equation were included in the pool of potential substrates. The final MetaGEM-DB dataset contains 330,432 non-redundant enzyme--metabolite interaction records, including 98,972 experimentally validated evidence entries and 231,460 high-confidence phylogenetic inference evidence entries.

\subsection{Data Sampling and De-redundant Splitting Strategy}
In biomacromolecular interaction prediction tasks, the ``extreme sparsity'' (class imbalance) of interaction relationships in nature and ``data leakage'' caused by sequence homology are two core challenges. To this end, we designed an extremely rigorous sampling and data splitting protocol.

\subsubsection{Hard Negative Mining}
To enhance the model's ability to distinguish structurally highly similar but biologically inactive ``decoy'' substrates, this study introduced a hard negative mining strategy. In constructing the candidate pool, negative samples were strictly limited to metabolites screened from the internal metabolite pool of MetaGEM-DB, ensuring that all candidates are real biological small molecules and have not undergone catalytic reactions with the target enzyme in the existing knowledge base. Unlike global random sampling, this strategy preferentially extracts metabolites with high chemical similarity to the true substrates (evaluated based on the Tanimoto coefficient of Morgan fingerprints\cite{bento2020open}) as hard negative samples. This mechanism forces the model to learn to distinguish subtle physicochemical structural differences near highly confounded decision boundaries. Meanwhile, to faithfully reflect the high sparsity of enzyme--substrate interaction networks in nature, the ratio of positive to negative samples during training was strictly fixed at 1:30.

\subsubsection{Homology-Aware Cluster-based Splitting}
To objectively evaluate the model's generalization ability to unseen enzyme families and fundamentally eliminate inflated performance caused by shallow sequence memorization, this study implemented rigorous data splitting at the clustering level. Specifically, the CD-Hit\cite{fu2012cd} algorithm was used to cluster all enzyme sequences under a preset sequence identity threshold. Subsequently, with clusters as the basic unit, the data were strictly divided into a training set and an independent test set at a ratio of 8:2. This homology-aware protocol ensures that there is no significant homologous overlap between the test set and the training set at the sequence level, thereby achieving a first-principles-level true evaluation.

\subsection{Baseline Models and Ablation Study Design}
Given that MetaGEM is the first to achieve end-to-end reconstruction from metabolomes directly to reverse inference of whole-genome metabolic networks, there is currently a lack of system-level predictive baselines in the field that can be directly compared. Therefore, our comparative validation strategy was decomposed into two progressive stages: at the core pairing module, we compared against existing advanced deep learning and machine learning predictors; at the system reconstruction stage, under a consistent network assembly configuration, we compared the topological properties and phenotypic prediction accuracy of the final generated metabolic networks.

\textbf{State-of-the-Art (SOTA) Baseline Models:}
To comprehensively benchmark the predictive performance of MetaGEM, this study selected three recent and representative enzyme--substrate interaction models for comparison:
(1) \textbf{ESP}\cite{kroll2023general}: a general feature representation model based on an improved Transformer architecture that enhances training by explicitly specifying randomly sampled small molecules as non-substrates;
(2) \textbf{VIPER}\cite{Campbell2024.06.21.599972} (Virtual Interaction Predictor for Enzyme Reactivity): a predictor focusing on improving generalization to unseen substrates, which systematically corrects annotation deficiencies in traditional datasets;
(3) \textbf{OmniESI}\cite{nie2025omniesi}: a unified framework based on progressive conditional deep learning that integrates prior knowledge such as enzyme catalytic specificity through conditional networks, thereby modulating deep features toward a catalysis-aware domain.

\textbf{Ablation Variants:}
To independently quantify the gain contributions of heterogeneous modality encoders and data quality in the MetaGEM architecture, we designed three modular variants:
(1) \textbf{UniESP$_{e}$}: replacing the molecular representation module in the ESP framework with the Uni-Mol2 encoder used in this study and retraining on the original ESP dataset to evaluate the independent gain of 3D geometric features;
(2) \textbf{UniESP$_{m}$}: maintaining the same network architecture as UniESP$_{e}$, but replacing the training set with the high-quality MetaGEM-DB constructed in this study to quantify the performance improvement brought by data curation;
(3) \textbf{MetaGEM$_{f}$}: as a core ablation variant of the complete framework, degrading its 3D geometric encoder into traditional 2D topological fingerprints (Morgan Fingerprints), aiming to verify the irreplaceability of three-dimensional spatial conformations in fine-grained substrate recognition.

\subsection{Evaluation Metrics and Validation Protocol}
This study constructed a two-level evaluation system covering microscopic interaction validation and macroscopic system functional validation.

\subsubsection{Evaluation of Enzyme--Metabolite Interaction Prediction}
On the independent de-homologized test set, we adopted standard binary classification metrics to quantify link prediction performance, including Accuracy, area under the precision--recall curve (AUPR), area under the receiver operating characteristic curve (AUROC), F1 score, and Matthews correlation coefficient (MCC). Given the severe inherent class imbalance in interaction data, the overall performance evaluation primarily considered AUPR and MCC, which are more sensitive to the minority class.

\subsubsection{Genome-Scale Metabolic Network Reconstruction and Functional Validation}
To validate the practical value of MetaGEM in generating system-level models with real biological fidelity, this study reconstructed the metabolic networks of three representative model organisms in a bottom-up manner: \textit{Escherichia coli}, \textit{Bacillus subtilis}, and \textit{Pseudomonas aeruginosa}. The quality of the reconstructed models was independently validated through the following two classical computational systems biology tasks:

\textbf{In Silico Phenotype Microarray Analysis:}
This task simulates cellular growth phenotypes under different single nutrient source conditions through flux balance analysis (FBA). All simulations were performed in an M9 minimal medium environment, where the maximum uptake rate of candidate carbon, nitrogen, phosphorus, and sulfur sources was uniformly constrained to $10 \text{ mmol gDW}^{-1} \text{ h}^{-1}$. In the testing workflow, the default nutrient sources in M9 medium (such as glucose and ammonium salts) were replaced one by one with the test compounds in the PM array plates. When the optimal flux solution with biomass generation as the objective function satisfied $\mu \geq 0.01 \text{ h}^{-1}$, the microorganism was determined to have the ability to utilize the corresponding substrate to sustain growth. The predicted binary growth states were then rigorously compared directly with literature-curated wet-lab phenotype microarray data. The experimental data were obtained from\cite{oh2007genome,feist2007genome,oberhardt2008genome,price2018mutant}.

\textbf{Gene Essentiality Analysis:}
This task evaluates the logical accuracy of the reconstructed network in predicting lethal phenotypes by simulating single-gene knockout in silico. The evaluation scope was strictly limited to the intersection of genes shared by the reconstructed network and the known experimental lethality dataset. According to the physiological characteristics of different species, the simulation environments were specifically set: \textit{E. coli} used M9 supplemented with glucose, \textit{P. aeruginosa} used M9 supplemented with succinate, whereas \textit{B. subtilis} used LB complex medium. Finally, based on the confusion matrix composed of predicted knockout results and experimental lethal phenotypes, the fidelity of the model's GPR mapping was comprehensively quantified using metrics such as Precision, Sensitivity, Specificity, F1 score, and Accuracy. The experimental data were obtained from\cite{oh2007genome,orth2011comprehensive,price2018mutant,turner2015essential,glass2006essential}.

\section{Results and Analysis}

\subsection{Benchmark Evaluation of Enzyme--Metabolite Interaction Prediction}
To rigorously evaluate the core predictive capability of MetaGEM at the microscopic molecular level, we conducted a comprehensive multidimensional benchmark test on a strictly de-homologized independent test set (sequence identity $< 40\%$), comparing it against three latest mainstream external baseline models (ESP, VIPER, and OmniESI) as well as three internal ablation variants.

\begin{figure*}
    \centering
    \includegraphics[width=0.95\linewidth]{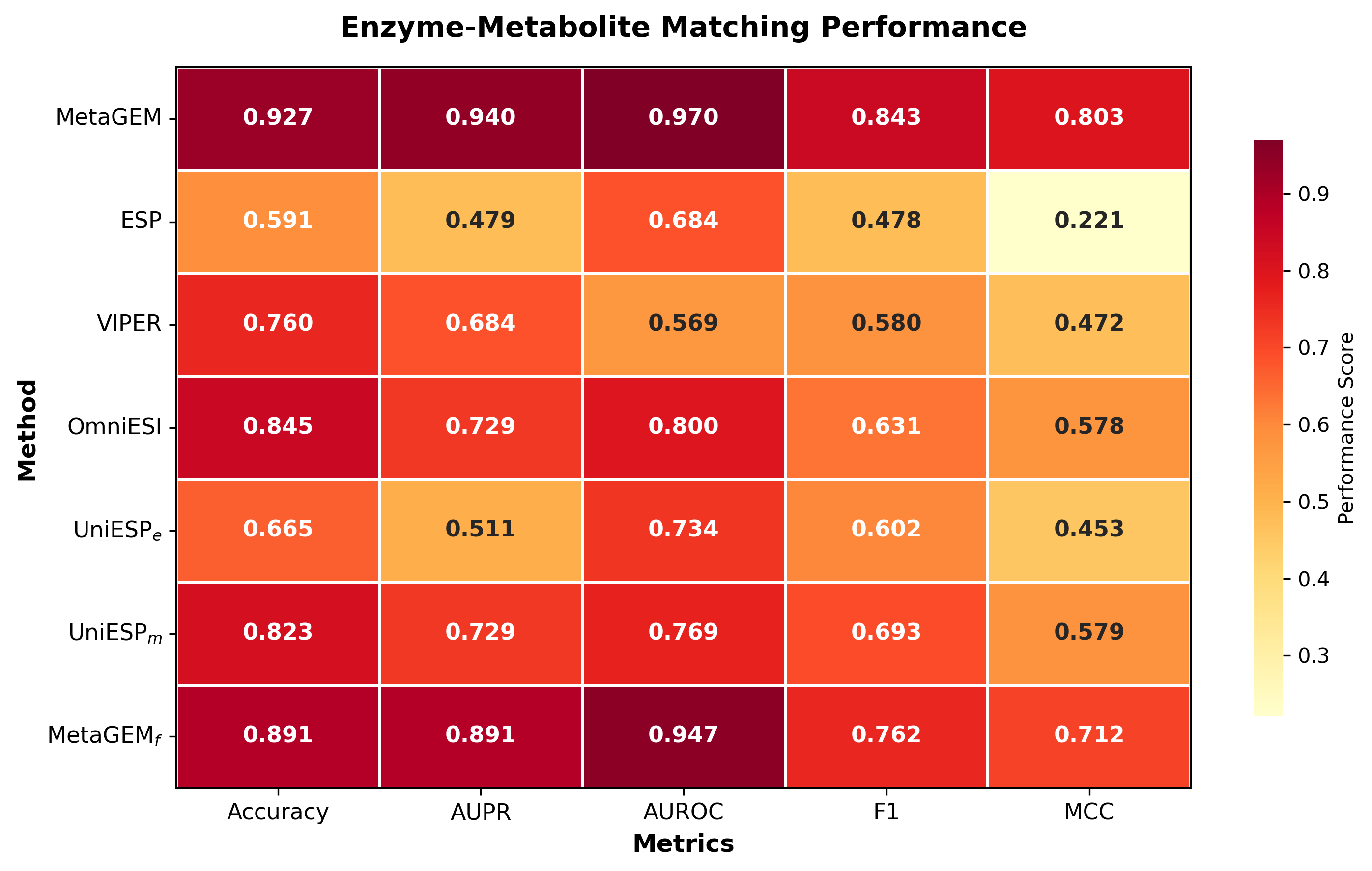}
    \caption{\textbf{Performance comparison of MetaGEM with SOTA baseline models and ablation variants on the enzyme--metabolite interaction (EMI) prediction task.} The heatmap in the figure shows the performance of seven models on five key metrics in the independent test set. Rows represent different evaluated models, and columns represent individual performance metrics. The color gradient from light yellow to dark red intuitively reflects the magnitude of prediction scores. MetaGEM, located in the first row, exhibits the darkest red across all evaluation metrics (achieving the best performance). In particular, on the core metrics AUPR and MCC, which measure the ability to handle extreme class imbalance, MetaGEM not only significantly outperformed the second-best baseline, but also, through direct comparison with the 2D fingerprint-based variant (MetaGEM$_{f}$), conclusively demonstrated the irreplaceability of introducing 3D spatial conformations for fine-grained substrate recognition.}
    \label{fig:EnzMetaMatching}
\end{figure*}

\subsubsection{Predictive Performance under Class Imbalance and Mechanistic Validation}
Biomacromolecular interaction data are inherently extremely sparse. Under a positive-to-negative sample ratio of 1:30, the conventional Accuracy metric is often severely dominated by the majority class (non-interacting background noise), thereby masking the model's true recognition ability on the minority class (real interacting pairs). The heatmap results in Figure~\ref{fig:EnzMetaMatching} strongly confirm this statistical trap: for example, although the second-best external baseline OmniESI maintained a relatively high superficial Accuracy (reaching 0.845), its MCC was only 0.578; meanwhile, the VIPER model, which was optimized for unseen substrates, had its core metric MCC for sparse recognition fall below 0.5 (only 0.472), and its AUROC was also extremely low (0.569). This fully exposes the discriminative limitations of traditional baseline models when dealing with extremely imbalanced data.

By contrast, the full MetaGEM established an absolute advantage across all key evaluation metrics, achieving an AUROC of 0.970 and an AUPR of 0.940. This all-round performance breakthrough indicates that the model can not only filter massive amounts of non-interacting background noise extremely effectively, but also precisely capture extremely scarce true catalytic relationships. More importantly, MetaGEM achieved a remarkable MCC of 0.803. Compared with traditional models adopting global random negative sampling strategies (such as ESP, whose MCC was only 0.221), MetaGEM successfully learned a robust decision boundary for handling subtle chemical structural differences by virtue of first-principles feature extraction and a rigorous hard negative mining mechanism.

In addition, the internal ablation experiments shown in the figure further reveal the underlying mechanistic sources of MetaGEM's performance leap. First, comparison between the two UniESP variants shows that, under the same network architecture, the model trained on the MetaGEM-DB constructed in this study (UniESP$_{m}$) consistently and significantly outperformed the model trained on the original dataset (UniESP$_{e}$) in both AUPR and F1 score. This result quantitatively highlights the underlying benefits brought by high-quality data curation and a rigorous homology-based splitting protocol. Second, when we forcibly downgraded MetaGEM's 3D geometric encoder to traditional two-dimensional topological fingerprints (i.e., the MetaGEM$_{f}$ variant), the model's core discriminative metric MCC exhibited a significant degradation of nearly 0.1 (from 0.803 to 0.712). This phenomenon provides direct experimental evidence that, when facing isomers or metabolites with similar two-dimensional topology but markedly different three-dimensional folded spaces, relying solely on planar fingerprints loses critical information on spatial steric hindrance and charge distribution; therefore, introducing 3D spatial geometric conformation as computational input is an absolutely necessary condition for achieving ``lock-and-key model''-level high-precision substrate pairing.

\subsubsection{Validation of Zero-Shot Generalization and ``Dark Matter'' Decoding Capability}
In natural microbial communities and vast metagenomic datasets, there exists a massive amount of ``metabolic dark matter'' lacking known homologous annotations. Traditional bioinformatics alignment-based methods often fail completely when handling these low-homology sequences. To rigorously evaluate MetaGEM's zero-shot generalization capability for decoding these unknown enzymes, we reconstructed the dataset and performed strict cross-validation according to different sequence identity thresholds (30\%, 50\%, 70\%, and 90\%).

\begin{table}[htbp]
    \centering
    \caption{\textbf{Anti-degradation performance evaluation of MetaGEM under different sequence identity thresholds.} The table shows the model performance on four progressively stricter de-homologized datasets. Even when the threshold was reduced to the extremely stringent 30\% (i.e., the enzymes in the test set are evolutionarily nearly orthogonal to those in the training set), the model's prediction metrics still remained remarkably stable, and even reached the optimal levels in AUPR and MCC. This counterintuitive anti-degradation phenomenon conclusively proves that the model learned underlying cross-species biochemical rules rather than simply memorizing shallow sequence features.}
    \label{tab:sequence_identity}
    \begin{tabular}{cccccc}
    \toprule
    \textbf{Sequence Identity} & \textbf{Accuracy} & \textbf{AUPR} & \textbf{AUROC} & \textbf{F1} & \textbf{MCC} \\
    \midrule
    30\% & \textbf{0.9234} & \textbf{0.9475} & \textbf{0.9749} & \textbf{0.8463} & \textbf{0.8059} \\
    50\% & 0.9175 & 0.9407 & 0.9656 & 0.8312 & 0.7905 \\
    70\% & 0.9161 & 0.9356 & 0.9674 & 0.8317 & 0.7883 \\
    90\% & 0.9193 & 0.9393 & 0.9699 & 0.8324 & 0.7921 \\
    \bottomrule
    \end{tabular}
\end{table}

Table~\ref{tab:sequence_identity} profoundly reveals MetaGEM's outstanding out-of-distribution (OOD) inference capability. In conventional deep learning models, as the sequence similarity between the training set and the test set decreases, the model often suffers a cliff-like drop in predictive performance due to ``representation degradation.'' However, the performance of MetaGEM breaks this convention: from the relatively relaxed 90\% homology down to the stringent 50\%, and even when entering the ``Twilight Zone'' of protein homology alignment ($<30\%$), the model's AUPR consistently remained above 0.935, and its MCC stayed robustly above 0.788.

More strikingly, under the most extreme 30\% sequence identity split (simulating the most realistic cold-start scenario for ``orphan genes''), the comprehensive discriminative performance of MetaGEM did not collapse, but instead achieved the globally optimal AUPR (0.9475) and MCC (0.8059). From the underlying logic of machine learning, under extremely low-homology clustering, the test set is forced to contain more challenging and diverse unseen protein fold types. The fact that the model's performance increased rather than decreased in this regime provides extremely strong mechanistic evidence that MetaGEM's dual-tower multimodal architecture has successfully transcended shallow memorization of local amino acid sequence motifs. What it relies on is the ``semantic--structure'' alignment between the deep macroscopic evolutionary semantics extracted by ESM and the microscopic three-dimensional geometric steric effects captured by Uni-Mol2.

This first-principles-based robust feature extraction endows the model with true zero-shot generalization capability. This means that even when facing a novel non-model biological chassis whose sequence has become completely unrecognizable due to prolonged independent evolution and for which no corresponding ``enzyme'' can be found in any traditional database, MetaGEM can still, by virtue of pure underlying physicochemical logic, discover its hidden metabolic functions with extremely high confidence. This capability completely removes the shackles of traditional homology annotation from metabolic reconstruction, providing a precise computational key for systematically illuminating the ``metabolic dark matter'' of life science.

\subsection{Deconstructing MetaGEM: Mechanistic Investigation and Feature Ablation}
To deeply dissect the intrinsic mechanisms underlying the superior performance of MetaGEM, we conducted a systematic ablation analysis of its core loss functions and feature extraction backbones.

\begin{table}[htbp]
\centering
\caption{\textbf{Ablation study of the effect of joint training strategies on model predictive performance.} The table compares model performance under standard binary cross-entropy (BCE), label smoothing (Label Smoothing BCE), weighted cross-entropy (Weighted BCE), focal loss (Focal), and the joint optimization strategy introducing hard negative mining (BCE + Contrastive Learning). After incorporating contrastive learning, the model achieved significant performance gains across all metrics, confirming the irreplaceability of the contrastive learning module in handling interference from highly similar molecules.}
\label{tab:loss_comparison}
\begin{tabular}{lccccc}
\toprule
\textbf{Loss Type} & \textbf{Accuracy} & \textbf{AUPR} & \textbf{AUROC} & \textbf{F1} & \textbf{MCC} \\
\midrule
BCE                          & 0.8753 & 0.8465 & 0.9340 & 0.7312 & 0.6670 \\
Label Smoothing BCE          & 0.8994 & 0.8737 & 0.9435 & 0.7927 & 0.7349 \\
Weighted BCE                 & 0.8799 & 0.8357 & 0.9227 & 0.7731 & 0.6917 \\
Focal                        & 0.9166 & 0.9147 & 0.9561 & 0.8261 & 0.7823 \\
BCE + Contrastive Learning   & \textbf{0.9266} & \textbf{0.9396} & \textbf{0.9701} & \textbf{0.8425} & \textbf{0.8033} \\
\bottomrule
\end{tabular}
\end{table}

\subsubsection{Contrastive Learning Reshapes the Topology of Latent Space}
The pervasive ``high-similarity interference'' in metabolite space is a core cause of false-positive predictions. We conducted a rigorous ablation analysis of the joint training strategy (Table~\ref{tab:loss_comparison}). The experiments not only tested the standard binary cross-entropy (BCE) loss, but also introduced label smoothing, weighted cross-entropy (Weighted BCE), and focal loss (Focal Loss) as strong baselines. The results show that merely performing weight adjustment or sample focusing on the binary classification objective is insufficient to fully overcome the performance bottleneck for minority-class samples (true interactions). For example, although Focal Loss improved MCC to 0.7823 by mining hard samples, after introducing the contrastive learning loss combined with hard negatives (BCE + Contrastive Learning), all key model metrics still achieved a globally significant improvement, with MCC reaching 0.8033. This fully confirms the irreplaceability of the contrastive learning module in handling highly deceptive structural analogs.

From the underlying mechanism of representation learning, this performance leap originates from the complete reshaping of the topological structure of the latent manifold by contrastive learning. Under the optimization of BCE alone and its variants, feature points of derivatives with high structural similarity are often entangled with each other in metric space, leading to blurred decision boundaries. Under the full MetaGEM framework, however, triplet margin loss forcibly constructs a highly clear ``clustering--repulsion'' topological structure: true enzyme--substrate pairs are tightly pulled together in space, whereas structurally highly similar ``decoy'' molecules (Hard Negatives) are explicitly pushed by a repulsive force field beyond the decision boundary (Margin $\alpha$). It is precisely this enforced separation in metric space that endows the model with first-principles-level fine-grained discriminative ability for distinguishing subtle chemical modifications (such as single-group substitution).

\begin{figure*}
    \centering
    \includegraphics[width=0.5\linewidth]{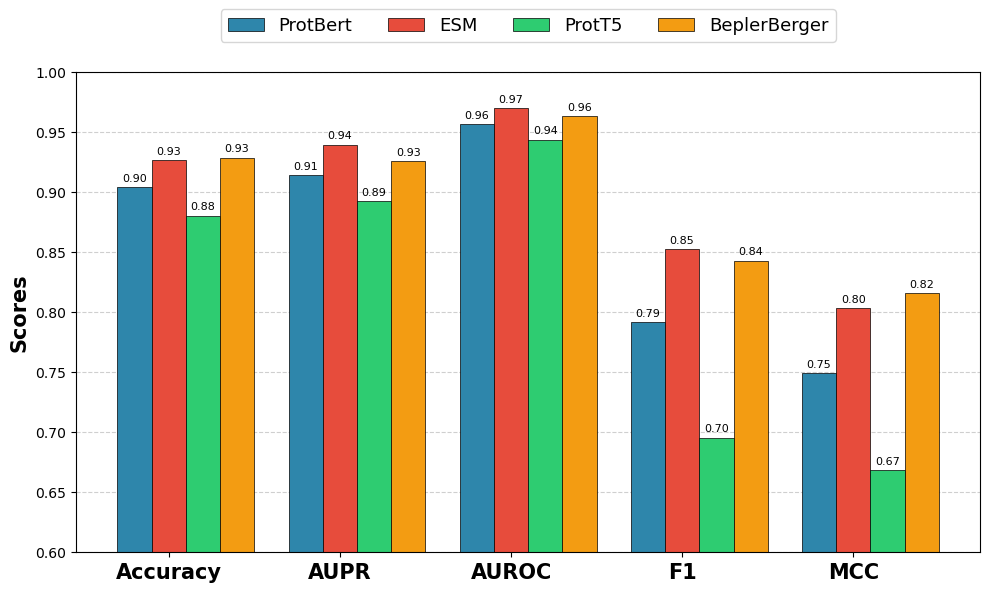}
    \caption{\textbf{Evaluation of the effects of different protein language models (PLMs) as sequence encoders.} The radar plot in the figure shows the performance polygons of four mainstream pretrained models (ProtBert, ESM, ProtT5, and BeplerBerger) on five key metrics. The fuller the outer contour, the stronger the overall performance. It can be observed that ESM (red line) and BeplerBerger (yellow line) form the two strongest outer boundaries, significantly outperforming ProtBert and ProtT5. Among them, BeplerBerger has a slight advantage in Accuracy and MCC, whereas ESM surpasses it on the core metrics measuring the ability to handle extreme imbalance (AUPR, AUROC, and F1), validating the strong representational power of deep evolutionary semantics in sparse molecular pairing.}
    \label{fig:plmablation}
\end{figure*}

\subsubsection{Feature Competition between Evolutionary Scale and Structural Priors}
The backbone ablation experiments on the protein encoder (Figure~\ref{fig:plmablation}) profoundly reveal the native impact of different pre-training paradigms on downstream fine-grained biochemical prediction performance. The radar plot results show clear performance stratification and reveal a feature competition between ``evolutionary scale'' and ``structural priors'' among the top-performing models.

Specifically, ProtT5 and ProtBert, which are trained on general corpora, show relatively limited performance, especially with obvious contraction in the F1 and MCC metrics when handling extremely imbalanced data (for example, ProtT5 achieves an MCC of only 0.6682). In contrast, ESM and BeplerBerger exhibit dominant performance. Notably, BeplerBerger even slightly surpasses ESM in Accuracy (0.9289) and the overall correlation coefficient MCC (0.8159). This phenomenon is highly consistent with its pretraining logic: BeplerBerger is pretrained in a supervised manner by explicitly leveraging protein three-dimensional contact maps and structural alignment information, and this strong ``structural prior'' endows the model with extremely high global classification boundary clarity.

However, in the real scenario of extremely sparse enzyme--metabolite interactions in nature, we place greater emphasis on AUPR and F1 score, which are more sensitive to minority-class positive samples. On these core metrics, ESM surpasses BeplerBerger (AUPR: 0.9396 vs 0.9260; F1: 0.8525 vs 0.8431). This reversal powerfully demonstrates a profound conclusion in computational biology: although ESM lacks explicit 3D structural supervision signals, through extremely deep unsupervised masked language modeling (MLM) on hundreds of millions of metagenomic sequences, it has successfully internalized implicit folding rules and active-site semantics shaped by billions of years of evolutionary pressure. This emergent ``deep evolutionary semantics'' is not only sufficient to rival manually annotated structural priors, but even exhibits higher robustness and specificity in filtering massive chemical decoys and capturing scarce catalytic relationships. This also provides conclusive experimental evidence for MetaGEM's final choice of the ESM architecture as the underlying foundation for exploring unknown metabolic ``dark matter.''

\begin{figure*}
    \centering
    \includegraphics[width=0.99\linewidth]{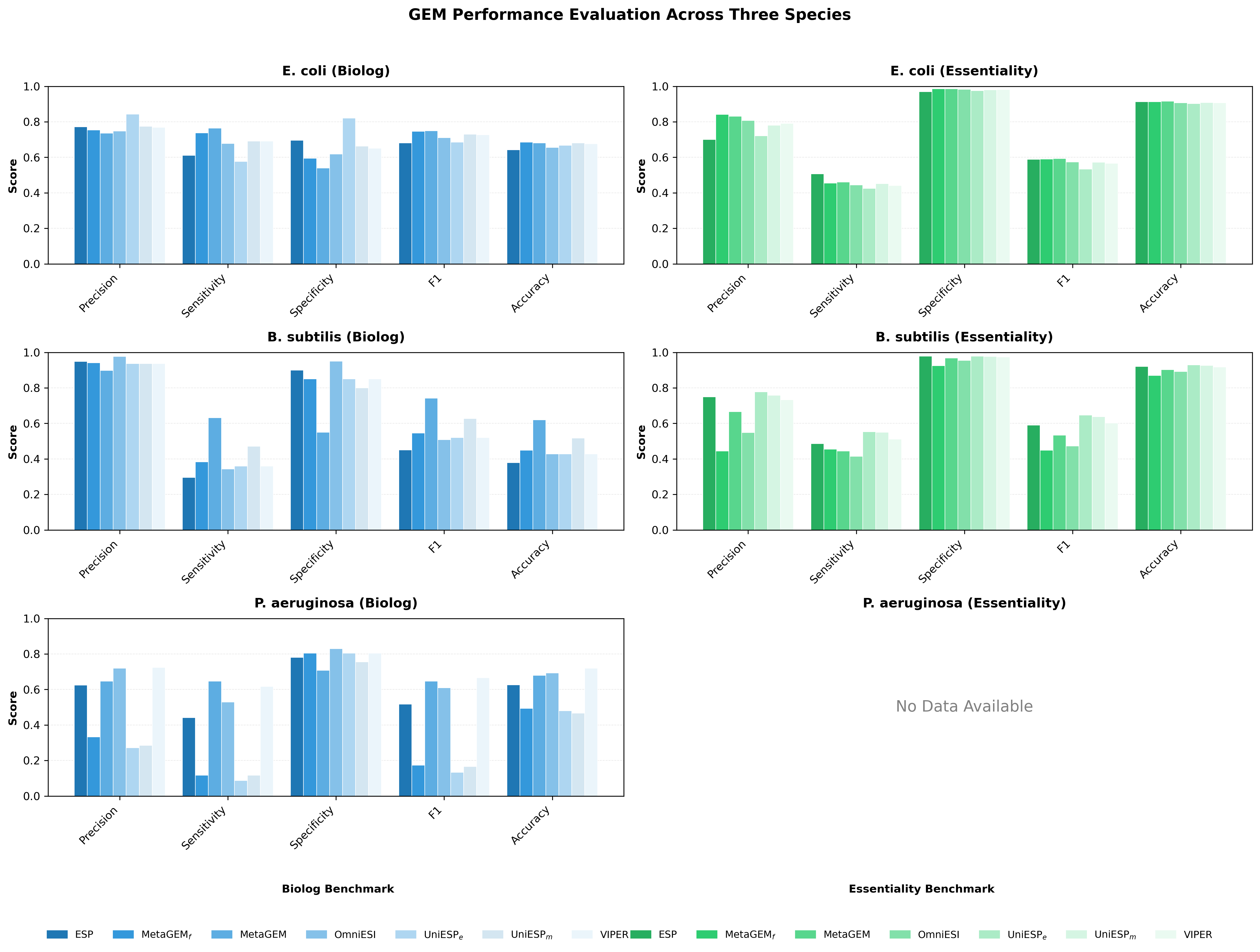}
    \caption{\textbf{Functional validation of reconstructed genome-scale metabolic models (GEMs) for three model organisms.} (Left column) In silico phenotype microarray (Biolog) analysis, showing the accuracy metrics for predicting cellular growth under different single nutrient sources. (Right column) Gene essentiality analysis, evaluating the accuracy of the models in predicting lethal gene knockouts. MetaGEM-driven models exhibit the highest consistency with experimental data in \textit{E. coli} and \textit{B. subtilis}. Note: data for the essentiality analysis of \textit{P. aeruginosa} are not shown because the input metabolomics data were extremely sparse, causing all reconstructed base models (wild type) to fail to grow in the specified medium (see the main text for details).}
    \label{fig:gem_barplot}
\end{figure*}

\subsection{System-Level Validation: Reverse Reconstruction of Genome-Scale Metabolic Networks}
The ultimate value of microscopic interaction prediction must be examined at the macroscopic scale of systems biology. Using the MetaGEM-driven end-to-end reconstruction pipeline, we reconstructed the metabolic networks of \textit{Escherichia coli}, \textit{Bacillus subtilis}, and \textit{Pseudomonas aeruginosa} in a bottom-up manner, and performed rigorous biological fidelity validation (Figure~\ref{fig:gem_barplot}).

\subsubsection{Phenotypic Simulation: Breaking through Network Disconnections and Connectivity Bottlenecks}
In the flux balance analysis (FBA)-based in silico phenotype microarray validation, the models reconstructed by MetaGEM showed excellent growth prediction performance under multiple nutrient environments (left column of Figure~\ref{fig:gem_barplot}). In particular, in \textit{B. subtilis} and \textit{P. aeruginosa}, which possess complex secondary metabolic capabilities, the sensitivity of baseline methods (such as ESP and VIPER) generally dropped to extremely low levels. Combined with the network topological scale (as shown in Table~\ref{tab:gem_stats_sub}), it can be observed that the networks generated by traditional methods contain significantly fewer total reactions. This indicates that they missed a large number of key catalytic reactions during inference, resulting in draft networks filled with severe gaps and dead-end metabolites, which are fundamentally incapable of sustaining basic growth phenotypes under most carbon sources. In contrast, MetaGEM, by virtue of its outstanding reverse inference capability, substantially improved the reaction coverage and phenotypic prediction sensitivity of the network. In the bottom-up reconstruction paradigm, this high-recall characteristic not only ensures that core metabolic pathways are not omitted, thereby providing a foundational guarantee for global network connectivity, but also avoids catastrophic compromise in specificity, demonstrating that the physicochemical constraints introduced in the subsequent assembly effectively prevent the persistent problem of over-gapfilling caused by purely mathematical programming.

\subsubsection{Gene Essentiality Analysis: System-Level Confirmation of High-Fidelity GPR Logic}
To further validate the accuracy of the underlying mapping logic of the model, we performed in silico single-gene knockout simulations (right column of Figure~\ref{fig:gem_barplot}). Gene essentiality prediction imposes dual and stringent requirements on the reconstructed network: the network must not only possess a complete biochemical reaction pool, but also require precisely mapped gene--protein--reaction (GPR) Boolean logic. The experimental results show that the models derived from MetaGEM comprehensively outperformed all baseline methods in essentiality prediction for \textit{E. coli} and \textit{B. subtilis}. This result provides conclusive system-level evidence that the high-confidence enzyme--metabolite pairings predicted by MetaGEM can be seamlessly and rigorously translated into system dynamic constraints. This not only proves the correctness of the model in reconstructing the physical topology of metabolic networks, but also confirms its high biological causal mapping fidelity in connecting static genetic information with dynamic phenotypic functions.

It is worth noting that in the essentiality analysis of \textit{P. aeruginosa} (lower right of Figure~\ref{fig:gem_barplot}), none of the models, including MetaGEM, produced valid evaluation data. In-depth computational systems diagnostics showed that this did not stem from inference errors of the model, but rather from the relative sparsity of the observed metabolite data used as system input for this species. In the ``bottom-up'' reconstruction paradigm, missing key metabolites leads the model to miss some ``hub enzymes'' in chassis metabolism. This structural deficiency directly causes all generated initial draft networks to fail to achieve basic wild-type flux growth in the specified simulated essentiality medium (M9 supplemented with succinate). Within the FBA framework, if the base model cannot produce biomass, any single-gene knockout simulation loses mathematical meaning.

This anomalous phenomenon instead objectively reveals an inherent boundary of ``phenotype-driven reverse engineering'': the completeness of network reconstruction is constrained by the breadth and quality of the input ``phenotypic snapshot (metabolome).'' Although MetaGEM had already restored biochemical connectivity far beyond baseline models to the greatest extent possible in the Biolog phenotype array of this species (with both phenotypic accuracy and F1 achieving the best performance, see the lower left of Figure~\ref{fig:gem_barplot}), the absence of a few core nodes still hindered specific whole-network dynamic simulations. This profound limitation also indirectly confirms an inherent boundary of phenotype-driven reverse engineering: the completeness of network reconstruction is fundamentally constrained by the breadth and quality of the input metabolomic snapshot. Achieving a truly comprehensive, blind-spot-free analysis of the virtual cell will therefore require complementary forward-reconstruction strategies that integrate genomic prior information alongside metabolomic evidence—a direction we identify as a critical avenue for future work (see Section 6.5).

\begin{table}
  \centering
  \caption{\textbf{Topological scale properties of reconstructed genome-scale metabolic models.} This table summarizes from left to right the absolute scale of the metabolic networks of three model organisms reconstructed by different methods, quantified by the total number of unique functional genes (\# Gene) and the total number of metabolic reactions (\# Reactions). Compared with all other baseline reconstruction pipelines and ablation variants, MetaGEM consistently recovered metabolic networks with the broadest reaction coverage. Notably, the increase in the number of reactions in MetaGEM is significantly greater than the increase in the number of genes, and this nonlinear amplification characteristic quantitatively confirms the model's absolute advantage in capturing promiscuous enzymes (multifunctional enzymes) and identifying metabolic hubs, thereby explaining from the perspective of physical topological scale its outstanding connectivity shown in phenotypic simulation.}
  \label{tab:gem_stats_sub}
  \begin{tabular}{lcccccc}
    \toprule
    & \multicolumn{2}{c}{\textit{E. coli}} & \multicolumn{2}{c}{\textit{B. subtilis}} & \multicolumn{2}{c}{\textit{P. aeruginosa}} \\
    \cmidrule(lr){2-3} \cmidrule(lr){4-5} \cmidrule(lr){6-7}
    \textbf{Method} & \# Gene & \# Reactions & \# Gene & \# Reactions & \# Gene & \# Reactions \\
    \midrule
    MetaGEM         & 1603 & 2482 & 1619 & 2580 & 1154 & 2131 \\
    ESP             & 1089 & 2238 & 1102 & 2304 &  770 & 1390 \\
    VIPER           & 1252 & 2330 & 1286 & 2519 &  887 & 1519 \\
    OmniESI         & 1348 & 2309 & 1376 & 2514 &  940 & 1576 \\
    UniESP$_e$      & 1118 & 2277 & 1212 & 2437 &  791 & 1401 \\
    UniESP$_m$      & 1168 & 2297 & 1229 & 2458 &  834 & 1412 \\
    MetaGEM$_{f}$   & 1450 & 2409 & 1489 & 2505 & 1001 & 1586 \\
    \bottomrule
  \end{tabular}
\end{table}

\subsubsection{Network Topological Characteristics: Exact Numerical Evidence for Promiscuous Enzyme Discovery Capability}
An in-depth quantitative analysis of the topological properties of the reconstructed networks (Table~\ref{tab:gem_stats_sub}) provides the most intuitive scale-based evidence for the superiority of MetaGEM. The data show that, in the reconstruction of all three model organisms, the network scale generated by MetaGEM (total number of reactions) is significantly larger than that of any baseline model. More importantly, the ``reaction increment'' and the ``gene increment'' in the model exhibit a highly nonlinear amplification trend. Taking \textit{P. aeruginosa} on the far right as an example, compared with the second-best OmniESI model, MetaGEM recruited only 214 additional functional genes (1154 vs 940), yet successfully recovered as many as 555 marginal biochemical reactions (2131 vs 1576). This reaction/gene gain ratio as high as 2.59 directly and powerfully demonstrates that the deep learning model greatly overcomes the narrowness of traditional sequence annotation methods, and successfully captures the widely existing promiscuous enzymes in organisms (i.e., the multifunctionality by which a single enzyme catalyzes multiple structurally similar substrates). This precise capture of enzyme multifunctionality effectively opens up the core hub nodes in metabolic networks, fundamentally alleviating the persistent problem of ``functional collapse'' frequently occurring in automated reconstruction, thereby making the generated mathematical models closer to the highly entangled complex panorama of real biological systems.

\subsection{Case Study: Breaking through Blind Spots of Homology Mapping to Precisely Reconstruct Key Metabolic Modules}

To specifically demonstrate the practical utility of MetaGEM in resolving unknown or easily overlooked metabolic pathways, we selected representative cases of differential gene essentiality prediction in two model organisms (\textit{B. subtilis} and \textit{E. coli}) for in-depth analysis. By comparing with mainstream automated reconstruction tools based on homology mapping (such as CarveMe), we intuitively clarify the core advantage of the bottom-up reconstruction paradigm in restoring the connectivity of key metabolic networks.

\subsubsection{\textit{Bacillus subtilis}: Opening the Entry Hub of the MEP Isoprenoid Precursor Biosynthesis Pathway}
In the gene essentiality benchmark test of \textit{Bacillus subtilis}, we observed that the \textit{dxs} gene (BSU24270) constitutes a typical empirical case of ``correctly predicted by MetaGEM (TP), but failed by baseline models (FN).'' This gene encodes 1-deoxy-D-xylulose-5-phosphate synthase (Dxs), which catalyzes the condensation of pyruvate and glyceraldehyde-3-phosphate to produce 1-deoxy-D-xylulose-5-phosphate (DXP). This biochemical reaction is the absolute entry step of the MEP (2-C-methyl-D-erythritol 4-phosphate) pathway, providing isoprenoid unit precursors such as IPP and DMAPP for \textit{B. subtilis} and many other microorganisms, thereby supporting the biosynthesis of quinones, lipid carriers, and other important membrane-related derivatives.

In real wet-lab data and the SubtiWiki database annotations, \textit{dxs} is explicitly annotated as an essential gene; multiple metabolic engineering studies have also consistently shown that Dxs is a key metabolic flux control point in this pathway, and its overexpression can significantly enhance isoprenoid product flux in \textit{B. subtilis}\cite{xue2015enhanced,wu2025metabolic}. However, during the execution of traditional sequence alignment-dependent reconstruction, this key reaction step was not included due to the inherent blind spots of homology thresholds, causing the model to lose the constraint of core precursor supply during single-gene deletion simulation, and thereby fail to correctly reproduce its lethal phenotype. In contrast, by virtue of its underlying high-precision enzyme--metabolite anchoring mechanism, MetaGEM successfully restored this entry reaction already at the draft stage. This case powerfully demonstrates that, compared with traditional reconstruction methods that are prone to missing single-point core nodes, MetaGEM can extremely robustly lock onto the key hubs of precursor pathways, ensuring that the physicochemical constraints of global metabolic flux are not disrupted.

\subsubsection{\textit{Escherichia coli}: Systematic Recovery of the Continuous Reaction Module of Histidine Biosynthesis}
If \textit{dxs} demonstrates MetaGEM's ability to capture a single key hub, then the histidine biosynthesis pathway in \textit{Escherichia coli} perfectly confirms its system-level recovery capability for continuous metabolic modules. In the evaluation, we observed an extremely significant clustered difference: the seven genes \textit{hisG}, \textit{hisD}, \textit{hisC}, \textit{hisB}, \textit{hisA}, \textit{hisF}, and \textit{hisI} all behaved as absolutely essential genes in experiments. MetaGEM achieved 100\% accurate prediction of the essentiality of these seven genes (all TP), whereas traditional baseline models collectively misclassified them as non-essential (all FN).

In-depth network topology diagnostics reveal that this is not an accidental random error. The enzymatic steps corresponding to these seven genes were completely and systematically restored in the network reconstructed by MetaGEM, whereas they were entirely missing in traditional models. From the biochemical logic perspective, HisG catalyzes the first reaction initiated by PRPP, HisA, HisF, and HisI participate in complex intermediate rearrangement and cyclization, HisB and HisC catalyze the subsequent dehydration and transamination steps, and HisD completes the terminal oxidation and ultimately produces histidine (consistent with KEGG pathway module annotations)\cite{winkler2009biosynthesis,murakami2015global}. The missing annotations of traditional methods did not occur at isolated single steps, but rather accumulated continuously along the entire pathway, ultimately causing the model to completely lose the biological constraints on this classical amino acid biosynthesis module. This case very clearly illustrates a core pain point: when homology mapping fails to stably recall continuous multistep reactions, traditional automated reconstruction tears apart complete biological pathways; in contrast, MetaGEM successfully ``assembles'' the complete connectivity of the pathway through layer-by-layer anchoring at the metabolite level in a bottom-up manner, thereby accurately reproducing the system-level gene essentiality phenotype.

\begin{figure*}
    \centering
    \includegraphics[width=0.95\linewidth]{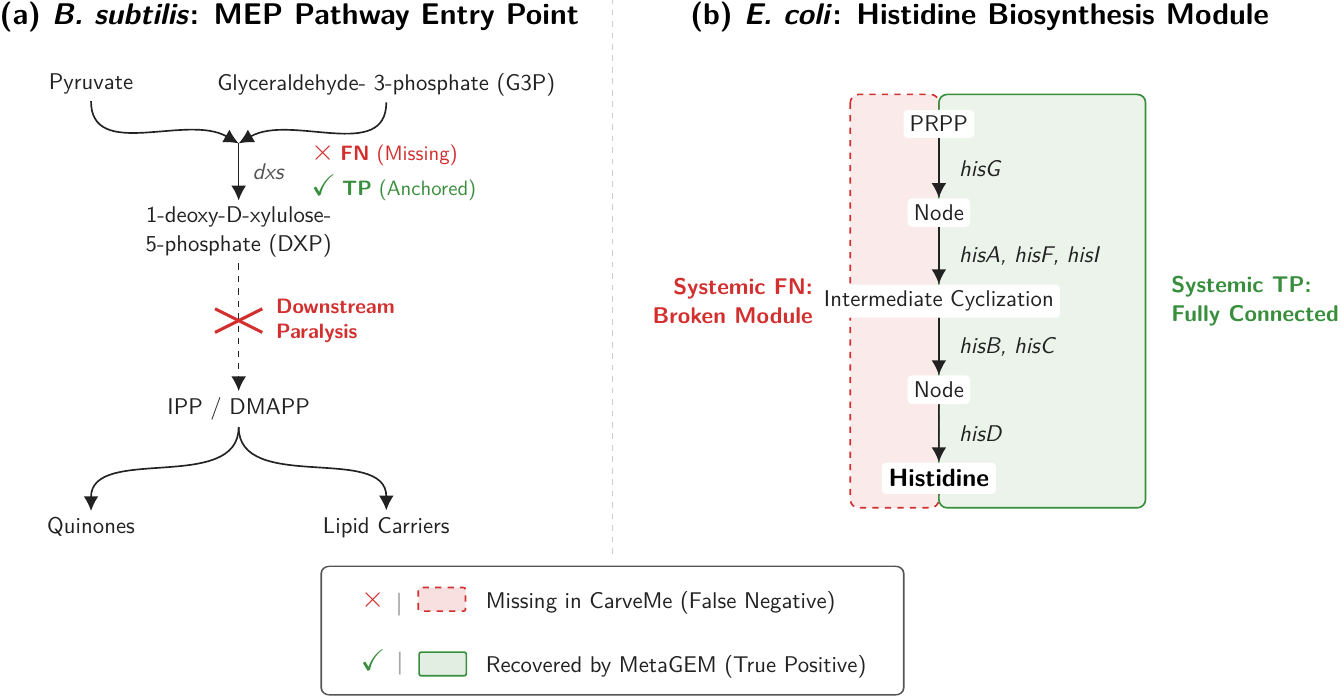}
    \caption{\textbf{Case analysis of MetaGEM in recovering key metabolic pathways and continuous biochemical modules.} (a) The MEP isoprenoid precursor biosynthesis pathway in \textit{B. subtilis}. Traditional homology mapping misses the entry hub enzyme Dxs, leading to downstream IPP precursor depletion; MetaGEM precisely anchors this reaction and correctly infers its essentiality. (b) The histidine biosynthesis module in \textit{E. coli}. Baseline models exhibit systematic missing annotation across the continuous seven-step cascade reaction (\textit{hisG} to \textit{hisD}), causing the entire metabolic branch to collapse; MetaGEM, through underlying physicochemical anchoring, completely assembles the connectivity topology from PRPP to histidine, achieving a 100\% hit rate for lethal phenotypes.}
    \label{fig:pathway_case_study}
\end{figure*}

\section{Discussion and Conclusion}

\subsection{Breaking the Top-Down Bottleneck: A Metabolite-Driven Reconstruction Paradigm}
The core objective of this study is to address a major challenge in the post-genomic era: to establish a computational framework that directly starts from the observed metabolite pool and reversely reconstructs functional genome-scale metabolic models (GEMs) in a ``bottom-up'' manner. For a long time, attempts to bypass sequence homology and directly map metabolites to macroscopic reaction networks have been constrained by severe combinatorial explosion and directional ambiguity. This study proposes that the only way to solve this ill-posed inverse problem is to identify a physical handle that anchors chemical potential to biological reality.

Based on this, we developed the MetaGEM framework, which strategically reduces the dimensionality of the macroscopic network inference problem into a microscopic molecular pairing task. By introducing the ``enzyme'' as the absolute anchor of network reconstruction (The Anchor), MetaGEM achieves a precise transformation from metabolite evidence to high-fidelity biochemical reactions. On a strict independent test set, MetaGEM achieved an AUROC of 0.9701 and an AUPR of 0.9396. These excellent microscopic prediction metrics are not the ultimate goal of this study, but rather provide a conclusive statistical guarantee that this deep learning engine can extract highly precise ``enzyme--metabolite'' interaction signals from massive background noise, thereby providing a seamless and contradiction-free high-quality component library for the downstream assembly of system-level GEMs.

\subsection{System-Level Validation: The Dual Triumph of Connectivity and High-Fidelity GPR}
Beyond pure machine learning prediction, the true value of MetaGEM lies in its outstanding performance in macroscopic system-level applications. In in silico phenotype microarray and gene essentiality analyses of \textit{E. coli}, \textit{B. subtilis}, and \textit{P. aeruginosa}, the models reconstructed by MetaGEM exhibited high biological fidelity.

It is particularly worth emphasizing that, in the ``bottom-up'' reconstruction paradigm, MetaGEM exhibited nonlinear network amplification characteristics and a keen ability to capture promiscuous enzymes. In the early stage of network draft construction, this high-recall (High Sensitivity) property constitutes a critically important computational strategic advantage: it maximally captures potential marginal biochemical reactions, effectively clears the hub nodes in metabolic networks, and fundamentally prevents the network fragmentation and ``functional collapse'' caused by missed predictions in traditional homology annotation-dependent methods. Subsequently, through precise gene--protein--reaction (GPR) logical mapping and mixed-integer linear programming (MILP), the model ensures whole-network connectivity while effectively suppressing over-gapfilling caused by purely mathematical constraints. These findings indicate that MetaGEM successfully transforms discrete metabolite space into a connected system-dynamics topology, faithfully reflecting the physicochemical constraints within cells.

\subsection{Deconstructing the Engine: Synergy of Multimodal First Principles and Contrastive Learning}
To support the high-precision reconstruction of the above macroscopic system, the precision of microscopic anchors is crucial. Ablation studies profoundly revealed the intrinsic mechanisms driving the underlying engine of MetaGEM. First, the dual-tower architecture integrating deep evolutionary semantics (ESM) and three-dimensional spatial conformations (Uni-Mol2) completely abandons dependence on shallow sequences and two-dimensional topological fingerprints, providing irreplaceable first-principles inputs for accurately simulating the ``lock-and-key'' matching mechanism.

Second, the joint training strategy fundamentally reshapes the topological structure of the feature manifold. To address the ``high-similarity interference'' in metabolic space that easily causes false positives in reconstruction, contrastive learning based on hard negative mining introduces a powerful inductive bias. It constructs a repulsive force field in latent space, forcibly pushing away structurally highly similar but biologically inactive ``decoy'' molecules. This mechanism effectively sharpens the decision boundary and endows the model with strong generalization capability for zero-shot inference toward non-model species and metabolic ``dark matter,'' thereby ensuring that the anchors of network reconstruction remain stable even in the face of extremely low-homology sequences.

\subsection{Conclusion and Perspective: Towards the Computational Foundation for AI-Driven Virtual Cells}
In summary, this study not only achieved SOTA performance on the specific task of enzyme--metabolite prediction, but more importantly, established a revolutionary new paradigm for GEM reconstruction. MetaGEM successfully opens the path from dynamic ``metabolomic snapshots'' to computable ``whole-genome network models,'' providing a completely new perspective for the deep analysis of complex biological systems.

In the grand vision of synthetic biology and biomanufacturing, the rational design of chassis cells highly depends on accurate metabolic mathematical models. However, the unresolved ``metabolic dark matter'' in massive numbers of non-model microorganisms has long constrained the boundaries of design. The bottom-up reconstruction capability provided by MetaGEM essentially offers a core computational foundation for the future metabolic integrated design of AI Virtual Cells. This breakthrough means that future digital twin cell systems will be able to directly read the host's metabolic potential phenotype, automatically evolve and complete their intrinsic biochemical dynamic models, thereby providing comprehensive and precise computational navigation for flux optimization of customized cell factories, mining of novel synthetic pathways, and dynamic network control.

\subsection{Limitations and Future Work}
Although MetaGEM pioneers a new route for reverse engineering of metabolic networks, this study still has certain limitations in moving toward the vision of fully digital twin cells, which also point the way for subsequent exploration in the field.

First, regarding the exploration of taxonomic boundaries, the high-quality training data currently relied upon by the model are mainly derived from species groups with relatively well-completed known annotations. This inherent limitation in pan-genomic breadth may, to some extent, affect the absolute inference accuracy of the model on extreme branches of the tree of life (such as deep-sea archaea or highly complex eukaryotes). In the future, it is urgent to extend the boundaries of data curation to a broader range of non-model organisms and metagenomic ``dark matter,'' so as to further enhance the phylogenetic universality of this deep learning framework.

Second, regarding full-dimensional automated simulation, although this study has successfully constructed an end-to-end pipeline from microscopic molecular pairing to macroscopic topological model assembly, challenges still remain in moving toward highly dynamic cellular simulation. When complex kinetic parameters and more refined whole-network thermodynamic constraints are introduced in the future, the current computational pipeline will still require deeper low-level algorithmic integration with advanced systems dynamics solvers, so as to overcome the limitations of static stoichiometry and ultimately achieve fully ``one-click'' dynamic metabolic flux simulation and prediction.

Finally, the ultimate confirmation of computational inference cannot be separated from the iterative loop of wet-lab experiments. Future work will focus on deeply integrating MetaGEM with high-throughput synthetic biology experimental platforms. By conducting targeted biochemical validation of newly predicted enzyme--substrate pairings and repaired secondary metabolic bypasses, we will truly establish a self-iterative flywheel of ``AI computational prediction--wet-lab validation--massive data feedback.'' This closed loop will not only continuously improve the fidelity of computational models, but also drive us closer to a comprehensive digital mapping of complex metabolic systems of life.

% References
\medskip

% Use the following code if you wish to generate your bibliography with BibTeX;
% replace the string "MSP-template" below with the name(s) of
% the BibTeX data base(s) you want to use.
% The resulting bibliography-output (the content of the .bbl file)
% must be pasted back into this file before submission.
% Please also include your BibTeX data base file(s) in your submission
% so that we can re-run BibTeX if necessary.
%
\bibliographystyle{MSP}
\bibliography{my_citation}

@article{thiele2010protocol,
  title={A protocol for generating a high-quality genome-scale metabolic reconstruction},
  author={Thiele, Ines and Palsson, Bernhard {\O}},
  journal={Nature protocols},
  volume={5},
  number={1},
  pages={93--121},
  year={2010},
  publisher={Nature Publishing Group UK London}
}

@article{lieven2020memote,
  title={MEMOTE for standardized genome-scale metabolic model testing},
  author={Lieven, Christian and Beber, Moritz E and Olivier, Brett G and Bergmann, Frank T and Ataman, Meric and Babaei, Parizad and Bartell, Jennifer A and Blank, Lars M and Chauhan, Siddharth and Correia, Kevin and others},
  journal={Nature biotechnology},
  volume={38},
  number={3},
  pages={272--276},
  year={2020},
  publisher={Nature Publishing Group US New York}
}

@incollection{baart2011genome,
  title={Genome-scale metabolic models: reconstruction and analysis},
  author={Baart, Gino JE and Martens, Dirk E},
  booktitle={Neisseria meningitidis: advanced methods and protocols},
  pages={107--126},
  year={2011},
  publisher={Springer}
}

@article{seaver2021modelseed,
  title={The ModelSEED Biochemistry Database for the integration of metabolic annotations and the reconstruction, comparison and analysis of metabolic models for plants, fungi and microbes},
  author={Seaver, Samuel MD and Liu, Filipe and Zhang, Qizhi and Jeffryes, James and Faria, Jos{\'e} P and Edirisinghe, Janaka N and Mundy, Michael and Chia, Nicholas and Noor, Elad and Beber, Moritz E and others},
  journal={Nucleic acids research},
  volume={49},
  number={D1},
  pages={D575--D588},
  year={2021},
  publisher={Oxford University Press}
}

@article{kharchenko2006identifying,
  title={Identifying metabolic enzymes with multiple types of association evidence},
  author={Kharchenko, Peter and Chen, Lifeng and Freund, Yoav and Vitkup, Dennis and Church, George M},
  journal={BMC bioinformatics},
  volume={7},
  number={1},
  pages={177},
  year={2006},
  publisher={Springer}
}

@article{machado2018fast,
  title={Fast automated reconstruction of genome-scale metabolic models for microbial species and communities},
  author={Machado, Daniel and Andrejev, Sergej and Tramontano, Melanie and Patil, Kiran Raosaheb},
  journal={Nucleic acids research},
  volume={46},
  number={15},
  pages={7542--7553},
  year={2018},
  publisher={Oxford University Press}
}

@article{mueller2013rapid,
  title={Rapid construction of metabolic models for a family of Cyanobacteria using a multiple source annotation workflow},
  author={Mueller, Thomas J and Berla, Bertram M and Pakrasi, Himadri B and Maranas, Costas D},
  journal={BMC systems biology},
  volume={7},
  number={1},
  pages={142},
  year={2013},
  publisher={Springer}
}

@article{dahal2016genome,
  title={Genome-scale modeling of thermophilic microorganisms},
  author={Dahal, Sanjeev and Poudel, Suresh and Thompson, R Adam},
  journal={Network Biology},
  pages={103--119},
  year={2016},
  publisher={Springer}
}

@article{de2024pan,
  title={pan-Draft: automated reconstruction of species-representative metabolic models from multiple genomes},
  author={De Bernardini, Nicola and Zampieri, Guido and Campanaro, Stefano and Zimmermann, Johannes and Waschina, Silvio and Treu, Laura},
  journal={Genome biology},
  volume={25},
  number={1},
  pages={280},
  year={2024},
  publisher={Springer}
}

@article{sorokina2014profiling,
  title={Profiling the orphan enzymes},
  author={Sorokina, Maria and Stam, Mark and M{\'e}digue, Claudine and Lespinet, Olivier and Vallenet, David},
  journal={Biology direct},
  volume={9},
  number={1},
  pages={10},
  year={2014},
  publisher={Springer}
}

@article{lobb2015remote,
  title={Remote homology and the functions of metagenomic dark matter},
  author={Lobb, Briallen and Kurtz, Daniel A and Moreno-Hagelsieb, Gabriel and Doxey, Andrew C},
  journal={Frontiers in genetics},
  volume={6},
  pages={234},
  year={2015},
  publisher={Frontiers Media SA}
}

@article{escudeiro2022functional,
  title={Functional characterization of prokaryotic dark matter: the road so far and what lies ahead},
  author={Escudeiro, Pedro and Henry, Christopher S and Dias, Ricardo PM},
  journal={Current Research in Microbial Sciences},
  volume={3},
  pages={100159},
  year={2022},
  publisher={Elsevier}
}

@article{palsson2026approaches,
  title={Approaches for accelerating microbial gene function discovery using artificial intelligence},
  author={Palsson, Bernhard O and Lee, Sang Yup and Kim, Gi Bae},
  journal={Nature Microbiology},
  pages={1--9},
  year={2026},
  publisher={Nature Publishing Group UK London}
}

@article{hsieh2024comparative,
  title={Comparative analysis of metabolic models of microbial communities reconstructed from automated tools and consensus approaches},
  author={Hsieh, Yunli Eric and Tandon, Kshitij and Verbruggen, Heroen and Nikoloski, Zoran},
  journal={npj Systems Biology and Applications},
  volume={10},
  number={1},
  pages={54},
  year={2024},
  publisher={Nature Publishing Group UK London}
}

@article{perez2021ultra,
  title={Ultra-high-performance liquid chromatography high-resolution mass spectrometry variants for metabolomics research},
  author={Perez de Souza, Leonardo and Alseekh, Saleh and Scossa, Federico and Fernie, Alisdair R},
  journal={Nature methods},
  volume={18},
  number={7},
  pages={733--746},
  year={2021},
  publisher={Nature Publishing Group US New York}
}

@article{zhang2024mass,
  title={Mass spectrometry imaging for spatially resolved multi-omics molecular mapping},
  author={Zhang, Hua and Lu, Kelly H and Ebbini, Malik and Huang, Penghsuan and Lu, Haiyan and Li, Lingjun},
  journal={npj Imaging},
  volume={2},
  number={1},
  pages={20},
  year={2024},
  publisher={Nature Publishing Group UK London}
}

@article{ali2022single,
  title={Single cell metabolism: current and future trends},
  author={Ali, Ahmed and Davidson, Shawn and Fraenkel, Ernest and Gilmore, Ian and Hankemeier, Thomas and Kirwan, Jennifer A and Lane, Andrew N and Lanekoff, Ingela and Larion, Mioara and McCall, Laura-Isobel and others},
  journal={Metabolomics},
  volume={18},
  number={10},
  pages={77},
  year={2022},
  publisher={Springer}
}

@article{gonccalves2021genome,
  title={Genome and metabolome: chance and necessity},
  author={Gon{\c{c}}alves, Emanuel and Frezza, Christian},
  journal={Genome Biology},
  volume={22},
  number={1},
  pages={276},
  year={2021},
  publisher={Springer}
}

@article{qiu2023small,
  title={Small molecule metabolites: discovery of biomarkers and therapeutic targets},
  author={Qiu, Shi and Cai, Ying and Yao, Hong and Lin, Chunsheng and Xie, Yiqiang and Tang, Songqi and Zhang, Aihua},
  journal={Signal transduction and targeted therapy},
  volume={8},
  number={1},
  pages={132},
  year={2023},
  publisher={Nature Publishing Group UK London}
}

@article{zhang2025dynamic,
  title={Dynamic single-cell metabolomics reveals cell-cell interaction between tumor cells and macrophages},
  author={Zhang, Yi and Shi, Mingying and Li, Mingxuan and Qin, Shaojie and Miao, Daiyu and Bai, Yu},
  journal={Nature communications},
  volume={16},
  number={1},
  pages={4582},
  year={2025},
  publisher={Nature Publishing Group UK London}
}

@article{buergel2022metabolomic,
  title={Metabolomic profiles predict individual multidisease outcomes},
  author={Buergel, Thore and Steinfeldt, Jakob and Ruyoga, Greg and Pietzner, Maik and Bizzarri, Daniele and Vojinovic, Dina and Upmeier zu Belzen, Julius and Loock, Lukas and Kittner, Paul and Christmann, Lara and others},
  journal={Nature medicine},
  volume={28},
  number={11},
  pages={2309--2320},
  year={2022},
  publisher={Nature Publishing Group US New York}
}

@article{nightingale2024metabolomic,
  title={Metabolomic and genomic prediction of common diseases in 700,217 participants in three national biobanks},
  journal={Nature communications},
  volume={15},
  number={1},
  pages={10092},
  year={2024},
  publisher={Nature Publishing Group UK London}
}

@article{moseley2013error,
  title={Error analysis and propagation in metabolomics data analysis},
  author={Moseley, Hunter NB},
  journal={Computational and structural biotechnology journal},
  volume={4},
  number={5},
  pages={e201301006},
  year={2013},
  publisher={Elsevier}
}

@article{engl2009inverse,
  title={Inverse problems in systems biology},
  author={Engl, Heinz W and Flamm, Christoph and K{\"u}gler, Philipp and Lu, James and M{\"u}ller, Stefan and Schuster, Peter},
  journal={Inverse Problems},
  volume={25},
  number={12},
  pages={123014},
  year={2009}
}

@article{li2023covrecon,
  title={COVRECON: automated integration of genome-and metabolome-scale network reconstruction and data-driven inverse modeling of metabolic interaction networks},
  author={Li, Jiahang and Waldherr, Steffen and Weckwerth, Wolfram},
  journal={Bioinformatics},
  volume={39},
  number={7},
  pages={btad397},
  year={2023},
  publisher={Oxford University Press}
}

@article{klamt2002combinatorial,
  title={Combinatorial complexity of pathway analysis in metabolic networks},
  author={Klamt, Steffen and Stelling, J{\"o}rg},
  journal={Molecular biology reports},
  volume={29},
  number={1},
  pages={233--236},
  year={2002},
  publisher={Springer}
}

@article{waller2020compartment,
  title={Compartment and hub definitions tune metabolic networks for metabolomic interpretations},
  author={Waller, T Cameron and Berg, Jordan A and Lex, Alexander and Chapman, Brian E and Rutter, Jared},
  journal={Gigascience},
  volume={9},
  number={1},
  pages={giz137},
  year={2020},
  publisher={Oxford University Press}
}

@article{blair2012can,
  title={What can causal networks tell us about metabolic pathways?},
  author={Blair, Rachael Hageman and Kliebenstein, Daniel J and Churchill, Gary A},
  journal={PLoS computational biology},
  volume={8},
  number={4},
  pages={e1002458},
  year={2012},
  publisher={Public Library of Science San Francisco, USA}
}

@article{krumholz2017thermodynamic,
  title={Thermodynamic constraints improve metabolic networks},
  author={Krumholz, Elias W and Libourel, Igor GL},
  journal={Biophysical journal},
  volume={113},
  number={3},
  pages={679--689},
  year={2017},
  publisher={Elsevier}
}

@article{sanchez2017improving,
  title={Improving the phenotype predictions of a yeast genome-scale metabolic model by incorporating enzymatic constraints},
  author={S{\'a}nchez, Benjam{\'\i}n J and Zhang, Cheng and Nilsson, Avlant and Lahtvee, Petri-Jaan and Kerkhoven, Eduard J and Nielsen, Jens},
  journal={Molecular systems biology},
  volume={13},
  number={8},
  pages={MSB167411},
  year={2017},
  publisher={Springer}
}

@article{yang2023improving,
  title={Improving pathway prediction accuracy of constraints-based metabolic network models by treating enzymes as microcompartments},
  author={Yang, Xue and Mao, Zhitao and Huang, Jianfeng and Wang, Ruoyu and Dong, Huaming and Zhang, Yanfei and Ma, Hongwu},
  journal={Synthetic and Systems Biotechnology},
  volume={8},
  number={4},
  pages={597--605},
  year={2023},
  publisher={Elsevier}
}

@article{li2022deep,
  title={Deep learning-based k cat prediction enables improved enzyme-constrained model reconstruction},
  author={Li, Feiran and Yuan, Le and Lu, Hongzhong and Li, Gang and Chen, Yu and Engqvist, Martin KM and Kerkhoven, Eduard J and Nielsen, Jens},
  journal={Nature catalysis},
  volume={5},
  number={8},
  pages={662--672},
  year={2022},
  publisher={Nature Publishing Group UK London}
}

@article{ryu2017framework,
  title={Framework and resource for more than 11,000 gene-transcript-protein-reaction associations in human metabolism},
  author={Ryu, Jae Yong and Kim, Hyun Uk and Lee, Sang Yup},
  journal={Proceedings of the National Academy of Sciences},
  volume={114},
  number={45},
  pages={E9740--E9749},
  year={2017},
  publisher={National Academy of Sciences}
}

@article{machado2016stoichiometric,
  title={Stoichiometric representation of gene--protein--reaction associations leverages constraint-based analysis from reaction to gene-level phenotype prediction},
  author={Machado, Daniel and Herrg{\aa}rd, Markus J and Rocha, Isabel},
  journal={PLoS computational biology},
  volume={12},
  number={10},
  pages={e1005140},
  year={2016},
  publisher={Public Library of Science San Francisco, CA USA}
}

@article{di2021gpruler,
  title={GPRuler: Metabolic gene-protein-reaction rules automatic reconstruction},
  author={Di Filippo, Marzia and Damiani, Chiara and Pescini, Dario},
  journal={PLoS computational biology},
  volume={17},
  number={11},
  pages={e1009550},
  year={2021},
  publisher={Public Library of Science San Francisco, CA USA}
}

@article{piazza2018map,
  title={A map of protein-metabolite interactions reveals principles of chemical communication},
  author={Piazza, Ilaria and Kochanowski, Karl and Cappelletti, Valentina and Fuhrer, Tobias and Noor, Elad and Sauer, Uwe and Picotti, Paola},
  journal={Cell},
  volume={172},
  number={1},
  pages={358--372},
  year={2018},
  publisher={Elsevier}
}

@article{kroll2023general,
  title={A general model to predict small molecule substrates of enzymes based on machine and deep learning},
  author={Kroll, Alexander and Ranjan, Sahasra and Engqvist, Martin KM and Lercher, Martin J},
  journal={Nature communications},
  volume={14},
  number={1},
  pages={2787},
  year={2023},
  publisher={Nature Publishing Group UK London}
}

@article{kroll2024multimodal,
  title={A multimodal Transformer Network for protein-small molecule interactions enhances predictions of kinase inhibition and enzyme-substrate relationships},
  author={Kroll, Alexander and Ranjan, Sahasra and Lercher, Martin J},
  journal={PLoS computational biology},
  volume={20},
  number={5},
  pages={e1012100},
  year={2024},
  publisher={Public Library of Science San Francisco, CA USA}
}

@article{nie2025omniesi,
  title={OmniESI: A unified framework for enzyme-substrate interaction prediction with progressive conditional deep learning},
  author={Nie, Zhiwei and Zhang, Hongyu and Jiang, Hao and Liu, Yutian and Huang, Xiansong and Xu, Fan and Fu, Jie and Ren, Zhixiang and Tian, Yonghong and Zhang, Wen-Bin and others},
  journal={arXiv preprint arXiv:2506.17963},
  year={2025}
}

@article{habibpour2024prediction,
  title={Prediction and integration of metabolite-protein interactions with genome-scale metabolic models},
  author={Habibpour, Mahdis and Razaghi-Moghadam, Zahra and Nikoloski, Zoran},
  journal={Metabolic Engineering},
  volume={82},
  pages={216--224},
  year={2024},
  publisher={Elsevier}
}

@article{salas2024machine,
  title={Machine learning to predict enzyme--substrate interactions in elucidation of synthesis pathways: a review},
  author={Salas-Nunez, Luis F and Barrera-Ocampo, Alvaro and Caicedo, Paola A and Cortes, Natalie and Osorio, Edison H and Villegas-Torres, Maria F and Gonz{\'a}lez Barrios, Andres F},
  journal={Metabolites},
  volume={14},
  number={3},
  pages={154},
  year={2024},
  publisher={MDPI}
}

@article{du2023fusing,
  title={Fusing 2D and 3D molecular graphs as unambiguous molecular descriptors for conformational and chiral stereoisomers},
  author={Du, Wenjie and Yang, Xiaoting and Wu, Di and Ma, FenFen and Zhang, Baicheng and Bao, Chaochao and Huo, Yaoyuan and Jiang, Jun and Chen, Xin and Wang, Yang},
  journal={Briefings in bioinformatics},
  volume={24},
  number={1},
  pages={bbac560},
  year={2023},
  publisher={Oxford University Press}
}

@article{orsi2024one,
  title={One chiral fingerprint to find them all},
  author={Orsi, Markus and Reymond, Jean-Louis},
  journal={Journal of cheminformatics},
  volume={16},
  number={1},
  pages={53},
  year={2024},
  publisher={Springer}
}

@article{tahil2024stereoisomers,
  title={Stereoisomers are not machine learning’s best friends},
  author={Tah{\i}l, Gökhan and Delorme, Fabien and Le Berre, Daniel and Monflier, {\'E}ric and Sayede, Adlane and Tilloy, S{\'e}bastien},
  journal={Journal of Chemical Information and Modeling},
  volume={64},
  number={14},
  pages={5451--5469},
  year={2024},
  publisher={ACS Publications}
}

@article{zhu2024metapredictor,
  title={MetaPredictor: in silico prediction of drug metabolites based on deep language models with prompt engineering},
  author={Zhu, Keyun and Huang, Mengting and Wang, Yimeng and Gu, Yaxin and Li, Weihua and Liu, Guixia and Tang, Yun},
  journal={Briefings in Bioinformatics},
  volume={25},
  number={5},
  pages={bbae374},
  year={2024},
  publisher={Oxford University Press}
}

@article{wang2020deep,
  title={Deep learning based drug metabolites prediction},
  author={Wang, Disha and Liu, Wenjun and Shen, Zihao and Jiang, Lei and Wang, Jie and Li, Shiliang and Li, Honglin},
  journal={Frontiers in Pharmacology},
  volume={10},
  pages={1586},
  year={2020},
  publisher={Frontiers Media SA}
}

@article{lin2022language,
  title={Language models of protein sequences at the scale of evolution enable accurate structure prediction},
  author={Lin, Zeming and Akin, Halil and Rao, Roshan and Hie, Brian and Zhu, Zhongkai and Lu, Wenting and Smetanin, Nikita and dos Santos Costa, Allan and Fazel-Zarandi, Maryam and Sercu, Tom and Candido, Sal and others},
  journal={bioRxiv},
  year={2022},
  publisher={Cold Spring Harbor Laboratory}
}

@inproceedings{
  zhou2023unimol,
  title={Uni-Mol: A Universal 3D Molecular Representation Learning Framework},
  author={Gengmo Zhou and Zhifeng Gao and Qiankun Ding and Hang Zheng and Hongteng Xu and Zhewei Wei and Linfeng Zhang and Guolin Ke},
  booktitle={The Eleventh International Conference on Learning Representations },
  year={2023},
  url={https://openreview.net/forum?id=6K2RM6wVqKu}
}

@article{schellenberger2010bigg,
  title={BiGG: a Biochemical Genetic and Genomic knowledgebase of large scale metabolic reconstructions},
  author={Schellenberger, Jan and Park, Junyoung O and Conrad, Tom M and Palsson, Bernhard {\O}},
  journal={BMC bioinformatics},
  volume={11},
  number={1},
  pages={213},
  year={2010},
  publisher={Springer}
}

@article{bansal2022rhea,
  title={Rhea, the reaction knowledgebase in 2022},
  author={Bansal, Parit and Morgat, Anne and Axelsen, Kristian B and Muthukrishnan, Venkatesh and Coudert, Elisabeth and Aimo, Lucila and Hyka-Nouspikel, Nevila and Gasteiger, Elisabeth and Kerhornou, Arnaud and Neto, Teresa Batista and others},
  journal={Nucleic acids research},
  volume={50},
  number={D1},
  pages={D693--D700},
  year={2022},
  publisher={Oxford University Press}
}

@article{uniprot2019uniprot,
  title={UniProt: a worldwide hub of protein knowledge},
  author={UniProt Consortium},
  journal={Nucleic acids research},
  volume={47},
  number={D1},
  pages={D506--D515},
  year={2019},
  publisher={Oxford University Press}
}

@article{gaulton2012chembl,
  title={ChEMBL: a large-scale bioactivity database for drug discovery},
  author={Gaulton, Anna and Bellis, Louisa J and Bento, A Patricia and Chambers, Jon and Davies, Mark and Hersey, Anne and Light, Yvonne and McGlinchey, Shaun and Michalovich, David and Al-Lazikani, Bissan and others},
  journal={Nucleic acids research},
  volume={40},
  number={D1},
  pages={D1100--D1107},
  year={2012},
  publisher={Oxford University Press}
}

@article{bento2020open,
  title={An open source chemical structure curation pipeline using RDKit},
  author={Bento, A Patr{\'\i}cia and Hersey, Anne and F{\'e}lix, Eloy and Landrum, Greg and Gaulton, Anna and Atkinson, Francis and Bellis, Louisa J and De Veij, Marleen and Leach, Andrew R},
  journal={Journal of cheminformatics},
  volume={12},
  number={1},
  pages={51},
  year={2020},
  publisher={Springer}
}

@article{fu2012cd,
  title={CD-HIT: accelerated for clustering the next-generation sequencing data},
  author={Fu, Limin and Niu, Beifang and Zhu, Zhengwei and Wu, Sitao and Li, Weizhong},
  journal={Bioinformatics},
  volume={28},
  number={23},
  pages={3150--3152},
  year={2012},
  publisher={Oxford University Press}
}

@article {Campbell2024.06.21.599972,
	author = {Campbell, Maxwell J.},
	title = {VIPER: A General Model for Prediction of Enzyme Substrates},
	elocation-id = {2024.06.21.599972},
	year = {2024},
	doi = {10.1101/2024.06.21.599972},
	publisher = {Cold Spring Harbor Laboratory},
	URL = {https://www.biorxiv.org/content/early/2024/12/16/2024.06.21.599972},
	eprint = {https://www.biorxiv.org/content/early/2024/12/16/2024.06.21.599972.full.pdf},
	journal = {bioRxiv}
}

@article{feist2007genome,
  title={A genome-scale metabolic reconstruction for Escherichia coli K-12 MG1655 that accounts for 1260 ORFs and thermodynamic information},
  author={Feist, Adam M and Henry, Christopher S and Reed, Jennifer L and Krummenacker, Markus and Joyce, Andrew R and Karp, Peter D and Broadbelt, Linda J and Hatzimanikatis, Vassily and Palsson, Bernhard {\O}},
  journal={Molecular systems biology},
  volume={3},
  pages={121},
  year={2007}
}

@article{oh2007genome,
  title={Genome-scale reconstruction of metabolic network in Bacillus subtilis based on high-throughput phenotyping and gene essentiality data},
  author={Oh, You-Kwan and Palsson, Bernhard O and Park, Sung M and Schilling, Christophe H and Mahadevan, Radhakrishnan},
  journal={Journal of Biological Chemistry},
  volume={282},
  number={39},
  pages={28791--28799},
  year={2007},
  publisher={Elsevier}
}

@misc{oberhardt2008genome,
  title={Genome-scale metabolic network analysis of the opportunistic pathogen Pseudomonas aeruginosa PAO1},
  author={Oberhardt, Matthew A and Pucha{\l}ka, Jacek and Fryer, Kimberly E and Martins dos Santos, V{\'\i}tor AP and Papin, Jason A},
  year={2008},
  publisher={American Society for Microbiology}
}

@article{price2018mutant,
  title={Mutant phenotypes for thousands of bacterial genes of unknown function},
  author={Price, Morgan N and Wetmore, Kelly M and Waters, R Jordan and Callaghan, Mark and Ray, Jayashree and Liu, Hualan and Kuehl, Jennifer V and Melnyk, Ryan A and Lamson, Jacob S and Suh, Yumi and others},
  journal={Nature},
  volume={557},
  number={7706},
  pages={503--509},
  year={2018},
  publisher={Nature Publishing Group UK London}
}

@article{orth2011comprehensive,
  title={A comprehensive genome-scale reconstruction of Escherichia coli metabolism—2011},
  author={Orth, Jeffrey D and Conrad, Tom M and Na, Jessica and Lerman, Joshua A and Nam, Hojung and Feist, Adam M and Palsson, Bernhard {\O}},
  journal={Molecular systems biology},
  volume={7},
  pages={535},
  year={2011}
}

@article{turner2015essential,
  title={Essential genome of Pseudomonas aeruginosa in cystic fibrosis sputum},
  author={Turner, Keith H and Wessel, Aimee K and Palmer, Gregory C and Murray, Justine L and Whiteley, Marvin},
  journal={Proceedings of the National Academy of Sciences},
  volume={112},
  number={13},
  pages={4110--4115},
  year={2015},
  publisher={National Academy of Sciences}
}

@article{glass2006essential,
  title={Essential genes of a minimal bacterium},
  author={Glass, John I and Assad-Garcia, Nacyra and Alperovich, Nina and Yooseph, Shibu and Lewis, Matthew R and Maruf, Mahir and Hutchison III, Clyde A and Smith, Hamilton O and Venter, J Craig},
  journal={Proceedings of the National Academy of Sciences},
  volume={103},
  number={2},
  pages={425--430},
  year={2006},
  publisher={National Academy of Sciences}
}

@article{xue2015enhanced,
  title={Enhanced C30 carotenoid production in Bacillus subtilis by systematic overexpression of MEP pathway genes},
  author={Xue, Dan and Abdallah, Ingy I and de Haan, Ilse EM and Sibbald, Mark JJB and Quax, Wim J},
  journal={Applied microbiology and biotechnology},
  volume={99},
  number={14},
  pages={5907--5915},
  year={2015},
  publisher={Springer}
}

@article{wu2025metabolic,
  title={Metabolic Engineering Strategy for Bacillus subtilis Producing MK-7},
  author={Wu, Shiying and Sun, Xiuwen and Fan, Tingwen and Lin, Fei and Chi, Yuan and Yang, Huaiyi and Zhao, Chunhui},
  journal={Foods},
  volume={14},
  number={23},
  pages={4150},
  year={2025},
  publisher={MDPI}
}

@article{winkler2009biosynthesis,
  title={Biosynthesis of histidine},
  author={Winkler, Malcolm E and Ramos-Monta{\~n}ez, Smirla},
  journal={EcoSal Plus},
  volume={3},
  number={2},
  pages={10--1128},
  year={2009},
  publisher={ASM Press Washington, DC}
}

@article{murakami2015global,
  title={Global coordination in adaptation to gene rewiring},
  author={Murakami, Yoshie and Matsumoto, Yuki and Tsuru, Saburo and Ying, Bei-Wen and Yomo, Tetsuya},
  journal={Nucleic acids research},
  volume={43},
  number={2},
  pages={1304--1316},
  year={2015},
  publisher={Oxford University Press}
}

\end{document}